\documentclass[]{aa}
\usepackage{graphicx}
\usepackage{txfonts}
\usepackage{natbib}
\bibpunct{(}{)}{;}{a}{}{,}
\begin{document}
\title{$\beta$~Cephei stars in the ASAS-3 data}
\subtitle{I. Long-term variations of periods and amplitudes}
\author{A.\,Pigulski\inst{1} \and G.\,Pojma\'nski\inst{2}}
\offprints{A.\,Pigulski}
\institute{Instytut Astronomiczny Uniwersytetu Wroc{\l}awskiego,
Kopernika 11, 51-622 Wroc{\l}aw, Poland\\
\email{ pigulski@astro.uni.wroc.pl} \and
Obserwatorium Astronomiczne Uniwersytetu Warszawskiego,
Al.~Ujazdowskie 4, 00-478 Warszawa\\
\email{gp@astrouw.edu.pl}}
\date{Received / Accepted}
\abstract{}
{We analyse $V$-filter ASAS-3 photometry of 41 known $\beta$~Cephei-type stars.
The ASAS-3 photometry is combined with the archival data, if available,
to determine long-term stability of periods and amplitudes of excited modes.}
{Frequencies of modes are derived by means of Fourier periodograms with consecutive
prewhitening.  The results are examined in the context of detection threshold.}
{We detected amplitude changes in three $\beta$~Cephei stars, BW Cru, V836 Cen,
and V348 Nor.  Period changes were found in KK~Vel and V836 Cen. Our analysis shows
that intrinsic period changes are more common among multiperiodic stars, apparently
because they are caused by some kind of mode interaction. In addition, we found new
modes for seven stars, and for ten others we provide new solutions or remove
ambiguities in the detected frequencies. One candidate hybrid $\beta$~Cephei/SPB
star, HD\,133823, is discovered.}
{}
\maketitle

\section{Introduction}
A group of main-sequence early B-type pulsating stars, known as $\beta$~Cephei stars,
has been studied for over a hundred years. The discovery of the mechanism driving
pulsations in these stars \citep{modz92, dzpa93, pamy99} advanced our
understanding of their nature and opened the possibility of addressing more complex
questions concerning their pulsations.  In particular, studies by means of
asteroseismology became possible.  The attractiveness of $\beta$~Cephei stars for
asteroseismology results from the fact that -- as pulsators -- they are neither too
simple nor too complex.  It is also one of the first groups of pulsating stars in
which nonradial pulsations were studied in detail, as nonradial pulsations are for
them the rule than the exception.  Moreover, ty\-pically less than a dozen modes
are observed in a single star which seems to be a sufficient number in seismic
modeling.

The properties of $\beta$~Cephei stars have recently been summarised in a review
paper of \citet{stha05}. Among other topics, these authors discussed the boundaries
of the instability strip of $\beta$~Cephei stars and pulsations in O-type stars.
There are many more interesting, yet still not well understood problems related
to $\beta$~Cephei stars, for example, their relation to other variable stars in
the upper main-sequence, influence of fast rotation on pulsations, dependence on
metallicity. One such problem, the long-term stability of periods and amplitudes
in these stars, is addressed in the present paper.

A reliable study of long-term changes of periods and/or amplitudes requires
long-time data with a good coverage. There are many $\beta$~Cephei stars that
show stable pulsation(s) on a time scale of decades. Only a handful are known for
their variation of period and/or amplitude \citep{jepi98,jerz99}. Sometimes, the
variation is fast enough to be detected using only observations from two
consecutive seasons. As possible causes of the period variation include
evolutionary effects and mode interaction, these changes merit being studied.

An all-sky survey covering several years of observations offers an excellent
opportunity to detect period and amplitude changes, especially if the data can
be combined with the archival ones. In the present paper we take advantage of
such an opportunity by analysing the photometry obtained within the ASAS-3
survey for all known $\beta$~Cephei stars in the magnitude range it covered.

\section{ASAS}
Photometric surveys, especially those that cover large parts of the sky, provide
an unprecedented chance of studying different classes of variable stars.  One
such survey, the All Sky Automated Survey \citep[ASAS,][and references therein]{pojm97,
pojm00, asas5}, already covers about 70\% of the whole sky. Nineteen new
large-amplitude $\beta$~Cephei stars were recently found using the published
ASAS catalogues \citep{pigu05,hand05i}.

Due to the method of selection, however, the published ASAS catalogues are
biased towards large-amplitude variables. Most of the nineteen $\beta$~Cephei
stars found in the ASAS-2 and ASAS-3 data by \citet{hand05i} and \citet[][hereafter
P05]{pigu05} have semi-amplitudes larger than 30~mmag, i.e., quite large for
$\beta$~Cephei stars. On the other hand, the typical detection threshold of
the ASAS data for stars brighter than 10~mag amounts to semi-amplitudes of
about 3--5 mmag. The 19 above-mentioned large-amplitude $\beta$~Cephei stars
represent therefore the tip of an iceberg and it was obvious that many more
such stars can be found in the ASAS data.

With this paper we start publication of the results of searching for
$\beta$~Cephei stars in the complete ASAS-3 database, i.e., among stars that
presently are not included in the published ASAS catalogues. Since many
known $\beta$~Cephei stars fall into the magnitude range of the ASAS-3
observations, we first publish the results of an analysis of the ASAS-3
photometry for known $\beta$~Cephei stars. This is the subject of the present
paper. Combining the ASAS-3 and archival data, we discuss here the frequency
contents of their pulsation spectra and the long-term stability of the
amplitudes and periods. The remaining two papers of the series will contain
the main results, namely the discovery of $\sim$280 $\beta$~Cephei stars. The first
part, 103 new $\beta$~Cephei stars, is presented in the accompanying paper
\citep[][hereafter Paper II]{pipo07} where we also discuss the presence of
low-frequency modes in these stars. The third paper of the series will contain data
for the remaining $\sim$180 $\beta$~Cephei stars found in the ASAS-3 data and
a discussion of the distribution of all known stars of this type in the Galaxy.

\section{The data}
The ASAS started in 1996 with the test phase, ASAS-1 and ASAS-2 \citep{pojm97, 
pojm98, pojm00}, which covered only selected areas in the southern sky and
equatorial regions. However, the first catalogue already included about 4000
bright variable stars \citep{pojm00}. A general description of the properties
of the ASAS-1 and ASAS-2 data and new automatic classification for this catalogue
was provided by \citet{eybl05}.

During the ongoing, third part of the survey, ASAS-3 \citep{pojm01}, started at the
end of 2000, the whole southern sky and partly the northern sky, up to declination
$+$28$\degr$, is monitored.  ASAS-3 provides photometry for about 15 $\times$ 10$^6$
stars in the range between 7 and 14~mag.  The accuracy of a single measurement
amounts to about 8 mmag for a star with $V\sim$ 8~mag, 13~mmag for a star with
$V\sim$ 10~mag, then increases rapidly.

By means of a magnitude-dispersion diagram, about 50\,000 variable stars were
selected using the ASAS-3 $V$-filter data. The stars were classified automatically
and included in the catalogue that is published \citep{asas1, asas2, asas3, asas4,
asas5}. The nineteen new $\beta$~Cephei stars mentioned above were found among
stars included in this catalogue.

\begin{figure}
\centering
\includegraphics[]{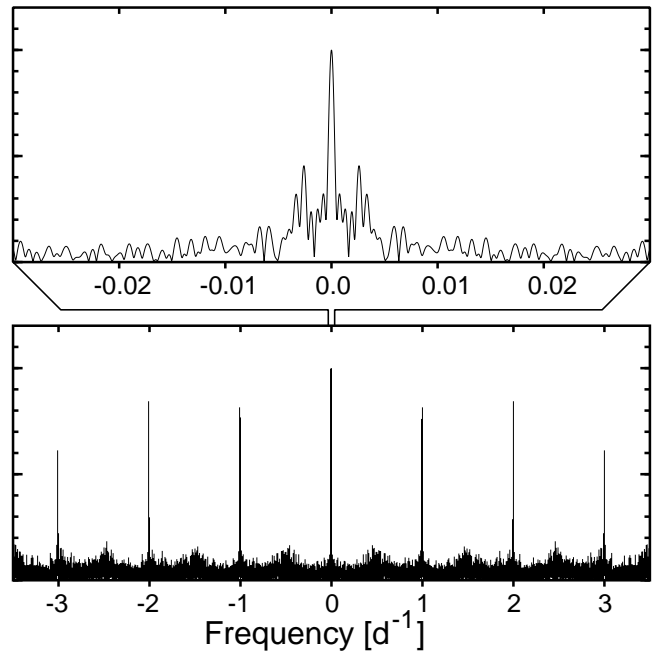}
\caption{Spectral window of a typical ASAS-3 data. Note that the yearly aliases at
$\pm$0.0027~d$^{-1}$ (top panel) are relatively low.  Ordinate is the amplitude in
arbitrary units.}
\label{sw}
\end{figure}

In the ASAS-3 data, typically a single measurement per night or two nights was
obtained for a given star.  This means that frequencies typical for $\beta$~Cephei
stars are located well above the Nyquist frequency.  Fortunately, the data are not
distributed ideally evenly in time. Consequently, the repeatable structure that for
evenly distributed data appears over the Nyquist frequency, does not occur in the
Fourier periodograms of the ASAS-3 data which were calculated in the interval between
0 and 40~d$^{-1}$.  The most severe problem with proper identification of frequencies
in the periodgrams of the ASAS-3 data was, as could be expected, related to the daily
aliases. This can be seen in Fig.~\ref{sw} where the spectral window of a typical
ASAS-3 data is shown.  On the other hand, yearly aliases are quite low in the ASAS-3
data (top panel of Fig.~\ref{sw}). This is a consequence of the fact that a given
field was observed for almost the whole year provided that it stood relatively high
above the horizon during the night.  This means that long gaps in the ASAS-3 data
occur only for stars located relatively close to the ecliptic and, due to location
of the site (Las Campanas Observatory, Chile), for stars with positive declinations.
The low yearly aliases in the periodograms of the ASAS-3 data are advantageous
for correct frequency identification.  While daily aliases can be easily removed
with a short follow-up multisite campaign, in order to remove the ambiguity in yearly
aliases, much longer campaigns would be required.

%
%
\begin{table*}
\begin{minipage}[t]{18cm}
\caption{Known $\beta$~Cephei stars with good ASAS-3 photometry. $N$ stands for the 
number of independent modes found in the ASAS-3 photometry, the remaining columns 
are self-explanatory.}
\label{bc-known}
\centering
\renewcommand{\footnoterule}{}
\begin{tabular}{rccccrrrll}
\hline \hline
 &  &  ASAS& Other & & \multicolumn{1}{c}{$V$} & $B-V$ & $U-B$ & &  \\
HD & CPD/BD & name & name & $N$ & [mag] & [mag] & [mag] & MK sp.~type
\footnote{References to the MK spectral types: [1]--\citet{garr77}, [2]--\citet{houk3}, [3]--\citet{houk2}, [4]--\citet{houk1}, [5]--\citet{evan05}, 
[6]--\citet{schi70}, [7]--\citet{hill70}, [8]--\citet{lepi01}, [9]--\citet{hill74}, [10]--\citet{morg53}, [11]--\citet{walb71}, [12]--\citet{kilk77}, 
[13]--\citet{fitz87}, [14]--\citet{whit63}, [15]--spectral type on the Harvard system, \citet{nest95}, [16]--\citet{houk4}.}
& Notes \\
\hline
59864 & $-$33$\degr$1536 & 073034$-$3405.4 & \object{V350 Pup} & 1 & 7.62 & $-$0.09 & $-$0.83 & B2\,II [1]& Ambiguity resolved\\
71913 & $-$34$\degr$2531 & 082843$-$3443.9 & \object{YZ Pyx} & 1 & 7.68 & $-$0.11 & $-$0.82 & B1/2 II [2] & \\
78616 & $-$44$\degr$3436 & 090742$-$4438.0 & \object{KK Vel} & 1 & 6.77 & $-$0.01 & $-$0.75 & B2 III [1] & Period changes\\
80383 & $-$52$\degr$2185 & 091731$-$5250.3 & \object{IL Vel} & 2 & 9.14 & $+$0.04 & $-$0.66 & B2 III [3] & \\
90288 & $-$56$\degr$3250 & 102357$-$5727.8 & \object{V433 Car} & 2 & 8.14 & $-$0.15 & $-$0.90 & B2 III/IV [4] &  \\
303067 & $-$57$\degr$3486 & 103530$-$5812.1 & \object{V401 Car}  & 1 & 9.54 & $+$0.03 & $-$0.76 & B1.5 III [5] & \\
--- & $-$57$\degr$3500 & 103541$-$5812.7 & \object{V403 Car} & 1 & 8.74 & 0.00 & $-$0.81& B1 III [5] & Ambiguity resolved\\
109885 & $-$70$\degr$1502 & 123920$-$7137.3 & \object{KZ Mus} & 3 & 9.02 & $+$0.13 & $-$0.66 & B2 III [4] & \\
--- & $-$59$\degr$4564 & 125358$-$6025.0 &  \object{BW Cru}  & 1 & 9.07 & $+$0.10 & $-$0.70 & B2 III [6] & Amplitude change\\
112481 & $-$49$\degr$5591 & 125736$-$4946.9 & \object{V856 Cen}& 1 & 8.35 & $+$0.05 & $-$0.76 & B2 II/III [3] &  \\
129929 & $-$36$\degr$6541 &  144626$-$3713.4 & \object{V836 Cen} & 5 & 8.09 & $-$0.18 & $-$0.87 & B3 V [7] & Ampl.~\& period changes  \\
145794 & $-$52$\degr$9416 & 161526$-$5255.3 & \object{V349 Nor}& 2 & 8.76 & $+$0.24 & $-$0.62 & B1 V [1] & \\
147985 & $-$43$\degr$7557 & 162657$-$4347.9 & \object{V348 Nor} & 2 & 7.95 & $+$0.14 & $-$0.64 & B1/2 II/III [3] & Amplitude change \\
156327 & $-$34$\degr$6800 & 171823$-$3424.5 & \object{V1035 Sco} & 3 & 9.35 & $+$0.62 & $-$0.11 & WC7 + B0 III [8] & New mode\\
156662 & $-$45$\degr$8479 & 172106$-$4558.9 & \object{V831 Ara}& 3 & 7.83 & $+$0.17 & $-$0.66 & B2 III [3]& Ambiguity resolved?\\
157485 & $-$26$\degr$5889 & 172435$-$2655.5 & \object{V2371 Oph}&2 & 9.07 & $+$0.56 & $-$0.30 & B1/2 Ib [2] & \\
164340 & $-$40$\degr$8357 & 180233$-$4005.2 & \object{NSV 24078}&4 & 9.28 & $-$0.14 & $-$0.95 & B0 III [9] & New mode\\
165812 & $-$22$\degr$6732 & 180845$-$2209.6 & \object{V4382 Sgr}&3 & 7.94 & $+$0.01 & $-$0.79 & B1.5 II [1] & New mode,\\
&&&&&&&&&ambiguity resolved\\
166540 & $-$16$\degr$4747 & 181148$-$1653.6 & \object{V4159 Sgr}&5 & 8.14 & $+$0.16 & $-$0.73 & B0.5 IV [10] & \\
180642 & $+$00$\degr$4159 & 191715$+$0103.6 & \object{V1449 Aql}& 1 & 8.27 & $+$0.22 & $-$0.66 & B1.5 II-III [11] & \\
203664 & $+$09$\degr$4973 & 212329$+$0955.9 & \object{SY Equ}&1 & 8.57 & $-$0.20 & $-$1.00 & B0.5 III(n) [11] & \\
--- & $-$19$\degr$8282 & 223738$-$1839.9 & \object{HN Aqr} &1 & 11.46 & --- & --- & B1 [12] & \\
\hline
--- & $-$62$\degr$2707 & 122213$-$6320.8 & \object{ALS 2653}&1 & 10.06 & $+$0.08 & $-$0.68 & B2 III [1] &\\
133823 & $-$65$\degr$2993 & 150955$-$6530.4 & & 4 & 9.62 & $+$0.05 & $-$0.66 & B2 IV [1] & New mode, hybrid object\\
--- & $-$50$\degr$9210 & 161858$-$5103.4 & \object{ALS 3547}&3 & 10.33 & $+$0.46 & $-$0.49 & B2 II [13] & New mode\\
328862 & $-$47$\degr$7861 & 164409$-$4719.1 & \object{ALS 3721}&3 & 10.13 & $+$0.27 & $-$0.51 & B0.5 III [14] &  \\
--- & $-$46$\degr$8213 & 164630$-$4701.2 & &1 & 10.86 & $+$0.56 & $-$0.39 & --- & New mode\\
328906 & --- & 164939$-$4431.7 & & 2 & 11.22 & --- & --- & B2 [15] & \\
152077 & $-$43$\degr$7731 & 165314$-$4345.0 & \object{ALS 3793} &3 & 9.08 & $+$0.29 & $-$0.58 & B1 II [1] & \\
152477 & $-$47$\degr$7958 & 165554$-$4808.8 & & 1 & 9.04 & $+$0.66 & $-$0.28 & B1 II [1] &\\
155336 & $-$32$\degr$4389 & 171218$-$3306.1 & \object{ALS 3961}& 3 & 9.46 & $+$0.25 & $-$0.59 & B1/2 Ib [2]& New solution\\
165582 & $-$34$\degr$7600 & 180808$-$3434.5 & \object{ALS 4668}& 4 & 9.39 & 0.00 & $-$0.80 & B1 II [1] & New solution\\
167743 & $-$15$\degr$4909 & 181716$-$1527.1 & & 3 & 9.69 & $+$0.35 & --- & B2 Ib [16] & \\
--- & --- & 182610$-$1704.3 & \object{ALS 5036} & 2 & 10.21 & $+$0.59 & $-$0.42 & --- & \\
--- & --- & 182617$-$1515.7 & \object{ALS 5040} & 3 & 10.75 & $+$0.47 & $-$0.51 & --- & New solution\\
--- & $-$14$\degr$5057 & 182726$-$1442.1 & & 1  & 9.97 & $+$0.51 & $-$0.31 & --- &\\
\hline
100495 & $-$62$\degr$2096 & 113318$-$6306.2 & \object{ALS 2386}&3 & 9.63 & $+$0.15 & $-$0.70 & B1 III [3] & New mode\\
--- & $-$61$\degr$3314 & 123748$-$6219.4 & \object{ALS 2714}&4 & 10.23 & $+$0.44 & $-$0.52 & --- & \\
--- & --- & 125319$-$6401.4 & \object{ALS 2798}& 3 & 11.65 & $+$0.92 & $-$0.15 & --- & \\
--- & --- & 130220$-$6328.4 & \object{ALS 2877}& 2 & 11.35 & --- & --- & --- & \\
191531 & $+$20$\degr$4449 &200940$+$2104.7 &  & 1 & 8.40 & $-$0.09 & --- & B0.5 III-IV [11] & \\
\hline
\end{tabular}
\end{minipage}
\end{table*}

The spatial resolution of the ASAS-3 data is defined by the detector scale which
amounts to about 14$^{\prime\prime}$ per pixel. For this reason, the ASAS-3 data are
not well suited for the study of stars in open clusters. Nevertheless, as we will
show below, some information on variability can be obtained from the ASAS-3 data
even for stars in clusters or other dense fields.  However, a strong contamination
by nearby stars and larger photometric errors can be expected in this case.

The ASAS-3 $V$-filter photometry analysed in this paper covers the interval between
the beginning of the project in 2000 and the end of February 2006.\footnote{The $V$
photometry for all 103 stars is available in electronic form at the CDS via anonymous 
ftp to cdsarc.u-strasbg.fr (130.79.128.5) or via
http://cdsweb.u-strasbg.fr/cgi-bin/qcat?J/A+A/??/??}

\section{Stars showing long-term period changes}
The recently published catalogue of $\beta$~Cephei stars \citep{stha05} contains
93 objects.  Of these, many fall within the magnitude range covered by the ASAS-3
data.  We present the results of analysis for 22 of them, i.e., all that have
reasonably good photometry in ASAS-3. The remaining are: (i) too bright, (ii) too
faint, (iii) north of declination $+$28$\degr$, (iv) in dense fields, e.g., open
clusters.  We also add a new analysis for 19 stars found with the ASAS data. Four
were found in the $I$-filter ASAS-2 data \citep{hand05i}. For the remaining 15,
some new $V$-filter observations were obtained since the time of publication of
the ASAS-3 data. General properties of all 41 stars under consideration are
summarised in Table \ref{bc-known}.  The results of sine-curve fits are presented
in Table \ref{bc-known-f}, \ref{bc-known-ap} and \ref{bc-known-gh} available as
online material. In addition, Fourier periodograms showing consecutive steps of
prewhitening are presented in Figs.~\ref{om1}--\ref{om-gh}, also available online.

\onltab{2}{%
\begin{table*}
\caption{Parameters of the sine-curve fits to the $V$ magnitudes of the known
$\beta$~Cephei-type stars listed in Table \ref{bc-known}. $N_{\rm obs}$ is the
number of observations. The initial epoch, $T_0$, equals 2450000.0.  The other
parameters are the following: $T_{\rm max}^i$ is the time of maximum light for
$i$th mode, $\sigma_{\rm res}$, standard deviation of the residuals, DT, detection
threshold defined as S/N = 4.}
\label{bc-known-f}
\begin{tabular}{rcclrcrr}
\hline \hline
  & & & \multicolumn{1}{c}{$f_i$} & \multicolumn{1}{c}{$A_i$} & $T_{\rm max}^i - T_0$ & 
\multicolumn{1}{c}{$\sigma_{\rm res}$} & \multicolumn{1}{c}{DT}\\
Star name &  Freq. & $N_{\rm obs}$ & \multicolumn{1}{c}{[d$^{-1}$]} & \multicolumn{1}{c}{[mmag]} & [d] 
& \multicolumn{1}{c}{[mmag]} & \multicolumn{1}{c}{[mmag]} \\
\hline
\object{V350 Pup} & $f_1$ & 529 & 4.23945(5) & 3.9(06) & 3016.0753(58) & 9.8 & 3.0\\
\object{YZ Pyx} & $f_1$ & 305 & 4.85955(2) & 17.1(09) & 2922.0226(17) & 11.2 & 4.5\\
\object{KK Vel} & $f_1$ & 282 & 4.63651(4) & 18.5(17) & 3238.1778(33) & 20.7 & 8.8\\
&2$f_1$ & & 9.27302 & 3.6(17) & 3238.0884(83) & &\\
\object{IL Vel} & $f_1$ & 457 & 5.45978(1) & 37.5(08) & 2930.4177(06) & 11.9 & 4.0 \\
& $f_2$ && 5.36325(1) & 34.4(08) & 2930.2800(07) &&\\
\object{V433 Car} & $f_1$ & 342 & 9.12945(3) & 8.7(10) & 2896.0042(18)& 12.0 & 4.6\\
& $f_2$ & & 8.31616(5) & 5.8(09) &2896.0604(31)  & & \\
\object{V401 Car} & $f_1$ & 302 & 5.92302(6) & 5.7(11) & 2954.4643(52) & 13.7 & 5.6\\
\object{V403 Car} & $f_1$ & 262 & 3.99009(4) & 16.6(21) & 3071.1564(50) & 24.0 & 10.4 \\
\object{KZ Mus} & $f_1$ & 632 & 5.86402(1) & 37.8(07) & 2817.6862(05) & 12.6 & 3.5\\
& $f_2$ & & 5.95070(1) & 15.9(07) & 2817.6471(12) & &\\
& $f_3$ & & 6.18750(2) & 10.9(07) & 2817.6239(17) & &\\
\object{BW Cru} & $f_1$ & 302 & 4.88458(4) & 10.3(12) & 2909.4986(38) & 14.5 & 5.9 \\
\object{V856 Cen} & $f_1$ & 273 & 3.92867(2) & 16.3(07) & 2819.7251(20) & 8.6 & 3.7 \\
\object{V836 Cen} & & 290 & \multicolumn{3}{c}{see Table \ref{v836cen-at}} & 15.8 & 6.6 \\
\object{V349 Nor} & $f_1$ & 387 & 6.25355(3) & 6.1(07) & 2961.9822(29) & 9.4 & 3.2\\
& $f_2$ & & 5.21317(6) & 3.9(07) &  2962.0157(53) &&\\
\object{V348 Nor} & $f_1$ & 354 & 6.90033(3) & 10.8(08) & 2784.2783(17) & 10.3 & 3.9 \\
& $f_2$ & & 7.55795(5) & 5.4(08) & 2784.2650(31) & &\\
\object{V1035 Sco} & $f_1$ & 678 & 6.84316(3) & 8.5(06) & 3050.3266(16) & 9.9 & 3.2\\
 & $f_2$ & & 7.46406(3) & 9.7(06) & 3050.3271(12) &&\\
 & $f_3$ & & 7.98473(4) & 5.8(06) & 3050.2980(18) &&\\
\object{V831 Ara} & $f_1$ & 251 & 6.30519(3) & 8.1(08) & 3012.4817(23) & 8.4 & 3.9\\
& $f_2$ & & 5.88907(5) & 4.9(08) & 3012.4255(42) & &\\
& $f_3$ & & 5.92138(5) & 4.4(08) & 3012.5105(45) & & \\
\object{V2371 Oph} & $f_1$ & 457 & 4.52105(2) & 23.2(09) & 3012.8319(13) & 12.8 & 4.8\\
 & $f_2$ & & 4.46677(2) & 14.3(09) & 3012.8161(21) & &\\
\object{NSV\,24078} & $f_1$ & 323 & 6.37773(1) & 27.7(08) &  2821.0650(08) & 10.4 & 4.1 \\
& $f_2$ & & 6.53875(1) & 22.0(08) & 2821.1148(09) & &\\
& $f_3$ & & 5.93043(4) & 6.5(08) & 2821.1187(34) & & \\
& $f_4$ & & 7.74246(6) & 4.8(08) & 2821.1315(36) & & \\
\object{V4382 Sgr} & $f_1$ & 713 & 5.68565(2) & 14.3(05) & 2968.9834(09) & 8.0 & 2.4\\
& $f_2$ & & 5.58499(2) & 11.2(05) & 2969.0571(11) & &\\
& $f_3$ & & 6.29401(5) & 3.3(05) & 2969.0984(33) & &\\
\object{V4159 Sgr} & $f_1$ & 581 & 4.25239(3) & 7.6(06) & 2946.1190(28) & 7.6 & 3.1\\
& $f_2$ & & 4.34200(3) & 9.5(05) &  2946.0906(18) & &\\
& $f_3$ & & 4.35640(3) & 6.3(05) & 2946.1130(28) & & \\
& $f_4$ & & 4.24964(3) & 7.8(06) & 2946.1301(28) & & \\
& $f_5$ & & 4.25765(6) & 3.9(05) & 2946.1956(46) & & \\
\object{V1449 Aql} & $f_1$ & 223 & 5.48694(1) & 37.4(10) & 3005.4092(08) & 10.4 & 4.9 \\
\object{SY Equ} & $f_1$ & 131 & 6.02877(2) & 30.1(11) & 2999.7624(10) & 9.0 & 5.6\\
\object{HN Aqr} & $f_1$ & 347 & 6.56525(3) & 20.7(22) & 2854.1722(25) & 28.3 & 10.8\\
\hline
\end{tabular}
\end{table*}
}%

\onltab{3}{%
\begin{table*}
\caption{Parameters of the sine-curve fits to the ASAS-3 $V$ magnitudes of 14
$\beta$~Cephei-type stars found by \citet{pigu05}. The headings are the same as in Table \ref{bc-known-f}.}
\label{bc-known-ap}
\begin{tabular}{rcclrlrr}
\hline \hline
  & & & \multicolumn{1}{c}{$f_i$} & \multicolumn{1}{c}{$A_i$} & $T_{\rm max}^i - T_0$ & 
\multicolumn{1}{c}{$\sigma_{\rm res}$} & \multicolumn{1}{c}{DT}\\
Star name &  Freq. & $N_{\rm obs}$ & \multicolumn{1}{c}{[d$^{-1}$]} & \multicolumn{1}{c}{[mmag]} & [d] 
& \multicolumn{1}{c}{[mmag]} & \multicolumn{1}{c}{[mmag]} \\
\hline
\object{CPD $-$62$\degr$2707} & $f_1$ & 458 & 7.058918(04) & 55.7(09) & 2845.4709(04) & 13.4 & 4.5\\
& 2$f_1$ & & 14.117836 & 9.6(09) &  2845.5446(11) & &\\
\object{HD 133823} & $f_1$ & 351 & 5.680441(06) & 53.7(11) & 2870.6894(05) & 13.6 & 5.3\\
& $f_2$ & & 0.50382(4) & 8.4(11) & 2870.429(39) & & \\
& $f_3$ & & 0.60818(5) & 7.3(11) & 2870.045(38) & & \\
& $f_4$ & & 0.59951(5) & 6.5(11) & 2870.819(43) & & \\
\object{CPD $-$50$\degr$9210} & $f_1$ & 365 &4.866871(10) & 39.7(13) & 2898.0067(11) & 17.1 & 6.0 \\
& $f_2$ & & 4.87937(2) & 20.3(13) & 2897.8218(21) & &\\
& $f_3$ & & 4.88575(6) & 6.8(13) &  2897.9157(62) & &\\
\object{HDE 328862} & $f_1$ & 329 & 4.948818(06) & 81.1(15) & 2882.8172(06) & 18.6 & 7.3\\
& $f_2$ & & 4.92461(3) & 15.7(15) & 2882.9150(31) & &\\
& $f_1$+$f_2$ & & 9.87343 & 10.4(15) & 2882.8640(23) & &\\
& $f_3$ & & 5.39896(5) & 9.3(15) & 2882.7266(46) & &\\
\object{CPD $-$46$\degr$8213} & $f_1$ & 301 & 4.460151(10) & 66.1(21) & 2923.6372(11) & 25.3 & 10.3\\
\object{HDE 328906} & $f_1$ & 341 & 5.630731(15) & 42.8(24) & 2902.1974(16) & 30.5 & 11.6\\
& $f_2$ & & 5.25868(5) & 14.3(23) & 2902.2620(51) & &\\
\object{HD 152077} & $f_1$ & 457 & 4.911492(07) & 48.8(11) & 2717.8316(08) & 16.7 & 5.6\\
& $f_2$ & & 4.851642(12) & 27.0(11) & 2717.7484(14) & &\\
& $f_3$ & & 4.88636(2) & 17.3(11) & 2717.8499(21) & &\\
& $f_1$+$f_2$ & & 9.76313 & 7.4(11) & 2717.7891(25) & &\\
\object{HD 152477} & $f_1$ & 395 & 3.773746(08) & 34.8(09) & 2891.5274(11) & 12.1 & 4.4\\
\object{HD 155336} & $f_1$ & 677 & 5.531722(08) & 46.7(08) & 3059.0981(05) & 14.7 & 5.3\\
& $f_2$ & & 5.45701(3) & 13.0(09) & 3059.0483(20) & &\\
& $f_3$ & & 5.06712(3) & 9.9(09) & 3058.9610(26) & &\\
\object{HD 165582} & $f_1$ & 776 & 4.747267(12) & 36.2(11) & 3398.4084(10) & 18.1 & 5.9\\
& $f_2$ & & 5.38372(3) & 17.2(11) & 3398.3529(19) & &\\
& $f_3$ & & 5.38766(4) & 11.8(11) & 3398.3701(28) & &\\
& $f_4$ & & 4.72497(3) & 11.1(11) & 3398.3376(34) & & \\
& $f_1$+$f_4$ & & 9.47223 & 8.9(10) & 3398.3792(18) & &\\
\object{HD 167743} & $f_1$ & 358 & 4.823723(08) & 41.2(10) & 2825.7459(08) & 13.5 & 5.1\\
& $f_2$ & & 5.09694(2) & 25.5(11) & 2825.8195(13) & &\\
& $f_3$ & & 4.97579(3) & 11.7(11) & 2825.7717(28) & &\\
\object{ALS 5036} & $f_1$ & 596 & 4.917347(07) & 56.5(11) & 2822.1409(06) & 18.2 & 6.1\\
& $f_2$ & & 4.91915(3) & 12.5(11) &  2822.0922(28)& &\\
\object{ALS 5040} & $f_1$ & 435 & 4.973834(11) & 47.5(16) & 2825.4432(10) & 21.8 & 7.5\\
& $f_2$ & & 5.07193(3) & 18.6(15) & 2825.5043(25) & &\\
& $f_3$ & & 5.52310(5) & 12.4(15) & 2825.3553(43) & &\\
\object{BD $-$14$\degr$5057} & $f_1$ & 385 & 4.163675(07) & 44.5(09) & 2823.1940(08) & 12.5 & 4.5\\
\hline
\end{tabular}
\end{table*}
}

\onltab{4}{%
\begin{table*}
\caption{Parameters of the sine-curve fits to the ASAS-3 $V$ magnitudes of five
$\beta$~Cephei-type stars found by \citet{hand05i}. The headings are the same as
in Table \ref{bc-known-f}.}
\label{bc-known-gh}
\begin{tabular}{rcclrcrr}
\hline \hline
  & & & \multicolumn{1}{c}{$f_i$} & \multicolumn{1}{c}{$A_i$} & $T_{\rm max}^i - T_0$ & 
\multicolumn{1}{c}{$\sigma_{\rm res}$} & \multicolumn{1}{c}{DT}\\
Star name &  Freq. & $N_{\rm obs}$ & \multicolumn{1}{c}{[d$^{-1}$]} & \multicolumn{1}{c}{[mmag]} & [d] 
& \multicolumn{1}{c}{[mmag]} & \multicolumn{1}{c}{[mmag]} \\
\hline
\object{HD 100495} & $f_1$ & 647 & 5.935344(08) & 26.9(08) & 2849.6155(08) & 13.9 & 3.7 \\
& $f_2$ & & 8.33112(5) & 4.4(08) & 2849.5807(35) & &\\
& $f_3$ & & 5.81785(6) & 4.1(08) & 2849.5591(53) & &\\
\object{CPD $-$61$\degr$3314} & $f_1$ & 379 & 4.598425(08) & 50.3(13) & 2915.1919(09) & 18.0 & 6.1\\
& $f_2$ & & 4.540054(11) & 31.0(13) & 2915.0874(15) & &\\
& $f_3$ & & 4.57789(2) & 25.4(13) & 2915.1198(18) & &\\
& $f_4$ & & 4.55636(2) & 20.3(13) & 2915.2895(23) & & \\
& $f_1$+$f_2$ & & 9.138479 & 16.0(13) &  2915.1315(14) & &\\
& $f_2$+$f_3$ & & 9.11794 & 8.6(13) & 2915.2070(27) & &\\
\object{ALS\,2798} & $f_1$ & 312 & 4.69547(2) & 50.1(24) & 2855.0179(16) & 28.5 & 11.4\\
& $f_2$ & & 4.68777(2) & 31.4(23) & 2854.9120(25) & &\\
& $f_3$ & & 4.44150(5) & 14.9(23) & 2854.9022(56) & &\\
\object{ALS\,2877} & $f_1$ & 315 & 5.28745(2) & 52.3(23) & 2852.2556(13) & 28.5 & 11.4\\
& $f_2$ & & 5.30272(3) & 25.4(23) & 2852.3951(27) & &\\
\object{HD 191531} & $f_1$ & 95 & 6.08590(3) & 27.5(16) & 3168.8461(15) & 10.9 & 7.9\\
\hline
\end{tabular}
\end{table*}
}

Combining the ASAS-3 and archival photometry, we have found period variations in
two stars, KK\,Vel and V836\,Cen. These findings are described in this section.
For V836\,Cen and two other stars, BW Cru and V348\,Nor, we also found long-term
variations of the amplitude (Sect.~5). The other individual stars, especially their
periods and amplitudes are commented on in Appendix A. For these stars, we did not
find long-term changes of periods and amplitudes either because they are stable in
the time covered by observations or the data are too scarce to allow that. However,
from the ASAS-3 observations we found new modes in seven stars; for ten others we
provide new solutions or resolve ambiguities in their frequencies. This is
commented on in the last column of Table \ref{bc-known}.

\onlfig{2}{
\begin{figure*}
\centering
\includegraphics[width=17cm]{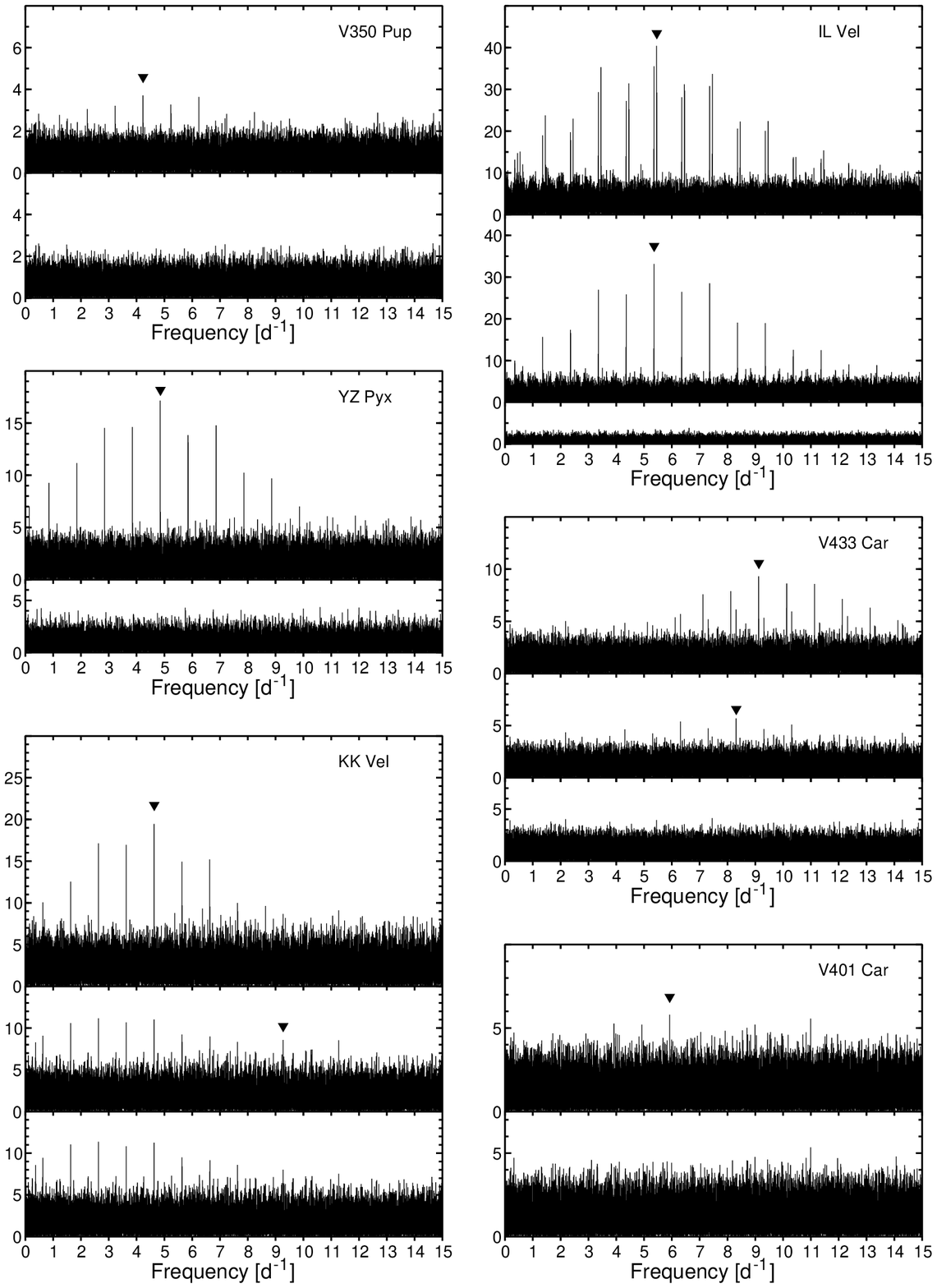}
\caption{Fourier periodograms of the ASAS-3 data of six $\beta$~Cephei-type stars
discussed in the paper: V350 Pup, YZ Pyx, KK Vel, IL Vel, V433~Car, and V401 Car.
The panels show periodograms after consecutive steps of prewhitening.  Frequencies
of the detected modes are indicated by inverted triangles.  Ordinate is the
semi-amplitude expressed in mmag.}
\label{om1}
\end{figure*}}

\onlfig{3}{
\begin{figure*}
\centering
\includegraphics[width=17cm]{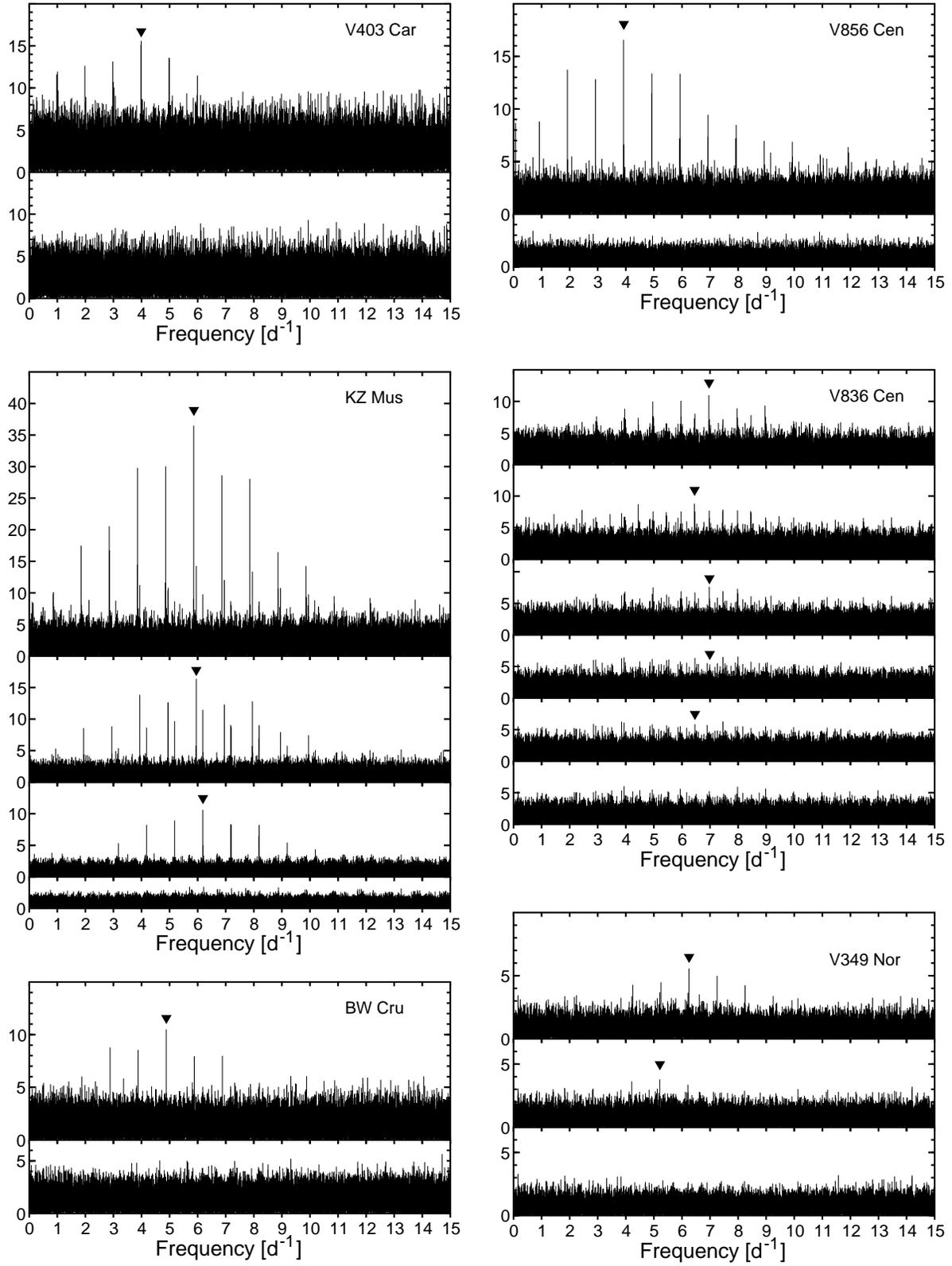}
\caption{The same as in Fig.~\ref{om1}, but for the next six $\beta$~Cephei stars:
V403 Car, KZ Mus, BW Cru, V856 Cen, V836 Cen, and V349 Nor.}
\label{om2}
\end{figure*}}

\onlfig{4}{
\begin{figure*}
\centering
\includegraphics[width=17cm]{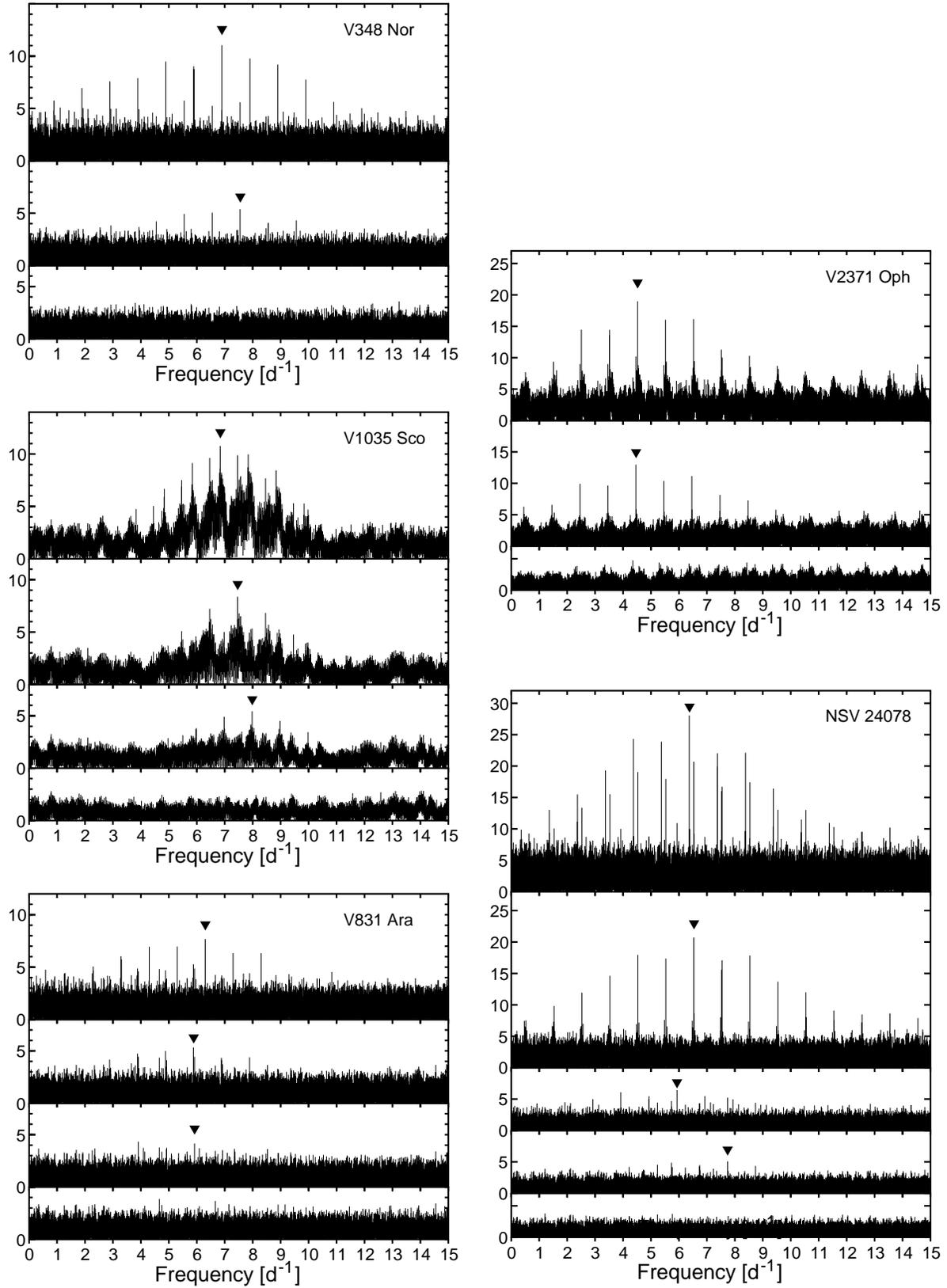}
\caption{The same as in Fig.~\ref{om1}, but for V348 Nor, V1035 Sco, V831 Ara,
V2371 Oph, and NSV 24078.}
\label{om3}
\end{figure*}}

\onlfig{5}{
\begin{figure*}
\centering
\includegraphics[width=17cm]{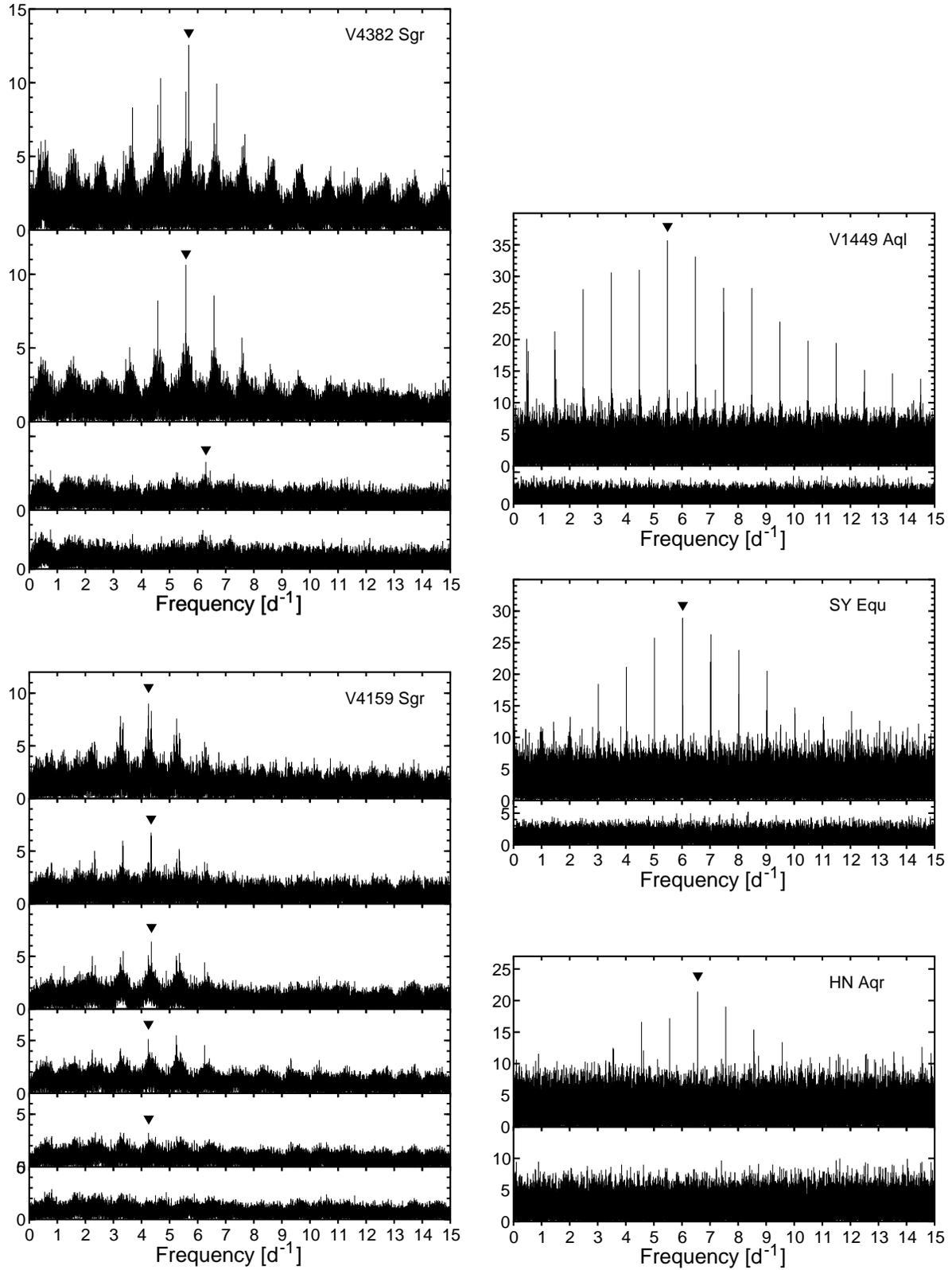}
\caption{The same as in Fig.~\ref{om1}, but for V4382 Sgr, V4159 Sgr, V1449 Aql,
SY Equ, and HN Aqr.}
\label{om4}
\end{figure*}}

\onlfig{6}{
\begin{figure*}
\centering
\includegraphics[width=17cm]{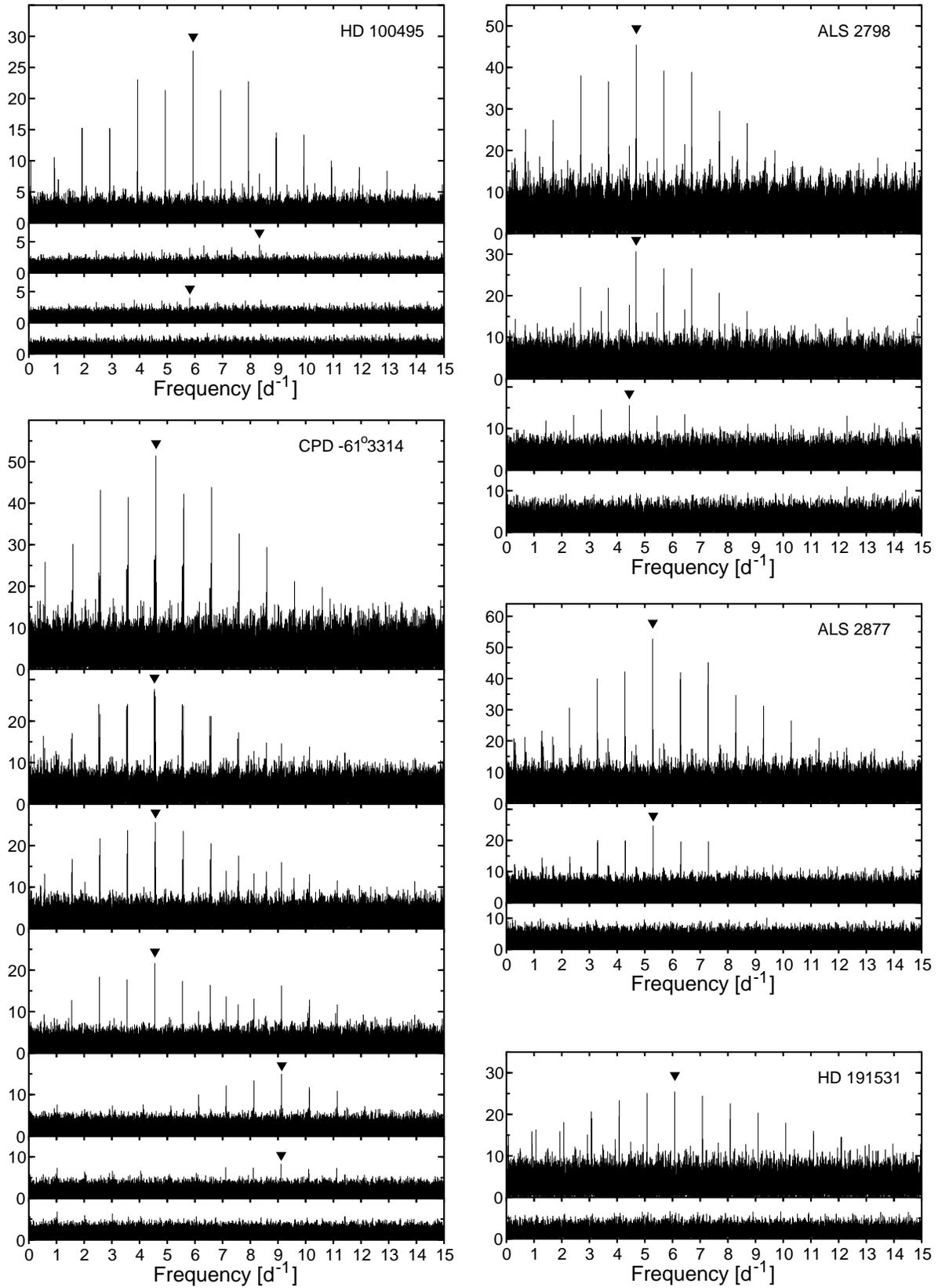}
\caption{The same as in Fig.~\ref{om1}, but for five stars discovered by
\citet{hand05i}: HD 100495, CPD\,$-$61$\degr$3314, ALS 2798, ALS 2877, and HD\,191531.}
\label{om-gh}
\end{figure*}}

\begin{figure}
\includegraphics[]{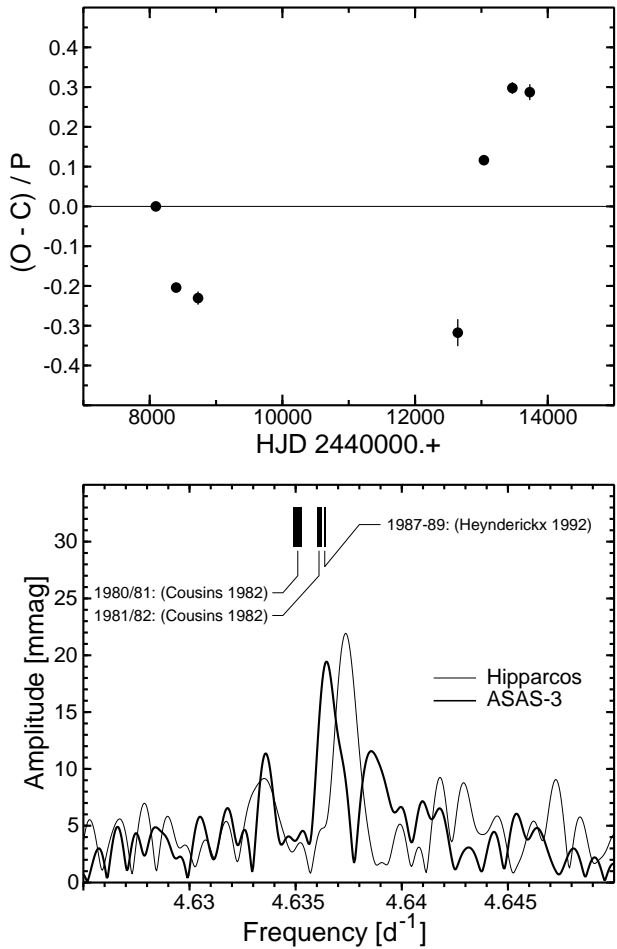}
\caption{{\it Top:} The O-C diagram for the main mode of KK Vel from the Hipparcos
and ASAS-3 data. {\it Bottom:} Fourier periodograms of the Hipparcos (thin line) and
ASAS-3 data (thick line) of KK Vel in the vicinity of the main peak. For comparison,
frequencies reported in the literature are shown with short vertical lines. The
widths of the lines correspond to the doubled r.m.s.~errors of the frequencies.}
\label{kkvel-pc}
\end{figure}

\subsection{\object{KK Vel} = \object{HD\,78616} = \object{HIP\,44790}}
The star was used as a standard for $UBV$ photometry in the Harvard E4 field.
Its variability was previously suspected by \citet{cost62}, but confirmed by
T.~Lloyd Evans in 1980 as reported by \citet{cous82}. \citet{cous82} indicated
also a non-sinusoidal shape of the light curve and possible period changes.
\citet{heyn92} analysed new Geneva and Walraven photometry of this star, also
detecting a single non-sinusoidal mode.  We recovered a single periodicity with
a harmonic both in the Hipparcos and the ASAS-3 data.  The presence of a harmonic
seems to be a little surprising in a star with not-so-large amplitude, but it
has to be remembered that KK Vel is a visual double.  The secondary is just
0$\farcs$3 distant and relatively bright, so that the observed amplitude is
reduced. After removing the contribution from the main signal and its harmonic
we still detect a strong signal at a frequency very close to the main frequency
(online Fig.~\ref{om1}). There is also a high peak close to the main frequency
in the periodogram of the residuals in the Hipparcos data.  We have split
Hipparcos and ASAS-3 data into seven subsets and derived amplitudes and times
of maximum light for each subset independently. The derived amplitudes are
constant within the errors, while the O-C diagram (Fig.~\ref{kkvel-pc})
clearly reveals changes of period. As mentioned above, the presence of period
changes in this star was already indicated by \citet{cous82}. He found that
the 1980--81 observations can be represented with a frequency of 4.6351 $\pm$
0.0002~d$^{-1}$, but 1981--82 data are best fitted with a frequency of 4.6361
$\pm$ 0.0001~d$^{-1}$.  The changes of period can be also seen directly in a
comparison of the periodograms of the Hipparcos and ASAS-3 data
(Fig.~\ref{kkvel-pc}).  Unfortunately, the existing data do not allow us to
trace these changes in detail.  It is worth noting, however, that the presence
of the visual companion of KK Vel may contribute to the apparent period changes
via the light-time effect like in $\beta$~Cep \citep{pibo92}, $\sigma$~Sco
\citep{pigu92}, and possibly BW Vul \citep{pigu93}.

\subsection{\object{V836 Cen}}
Since the amplitude changes in V836~Cen are better pronounced than period
changes, a detailed description of the analysis of the ASAS and archival data
for this star is given in Sect.~5.1. Of the six modes observed in V836~Cen
only one seems to be relatively stable (Fig.~\ref{v836cen-aoc}). The remaining
show cyclic or even more complex behaviour. A similar character of period
changes is observed in 12 Lac \citep{pigu94}. Both stars are similar in the
sense that they have several modes with considerable amplitudes. There is no
simple explanation of the period changes seen in V836~Cen, we can only
speculate that they might be caused by some kind of mode interaction. This is
discussed further in Sect.~6.

\section{Stars with long-term amplitude changes}
The most interesting star showing detectable amplitude changes is V836~Cen.
In addition, we were able to detect amplitude changes in two other
stars, BW Cru and V348 Nor.

\subsection{\object{V836 Cen} = \object{HD\,129929} = \object{HIP\,72241}}
The variability of HD\,129929 was discovered by \citet[][see also 
\citet{ruba82}]{rufe81}. \citet{waru83} found three periodic terms in the
photometry of this star. \citet{heyn92} also found three significant modes,
but their periods did not agree well with those of \citet{waru83}.
\citet[][hereafter A04]{aert04} combined data obtained by previous
investigators and new ones. The dataset analysed by these authors consisted
of 21 years of observations in the seven-band Geneva photometric system. Their
analysis yielded six independent modes including equally-spaced triplet around
6.98~d$^{-1}$ and a doublet with the same spacing around 6.46~d$^{-1}$. The
possibility of the presence of additional low-amplitude modes was also
indicated by A04.

The first mode we detected in the ASAS-3 data was the $f_6$ mode in the
\citet[][]{aert04} notation (we will use this notation for V836 Cen throughout
this section), i.e., the mode which had the smallest amplitude of the six modes
that were found in the Geneva data. The next two modes we detected were $f_3$
and a mode with a frequency of 6.97767~d$^{-1}$. This is very close to $f_2$ =
6.978305~d$^{-1}$ (A04), but is significantly different from it.  We temporarily
adopt 6.97767~d$^{-1}$ as the true frequency.  The periodogram with three modes
subtracted revealed daily aliases of $f_1$ and $f_4$. However, both were at the
limit of detection as the signal-to-noise (S/N) ratio for these peaks amounted
to 4.0. The $f_5$ mode was not detected at all.
\begin{figure*}
\sidecaption
\includegraphics[width=12cm]{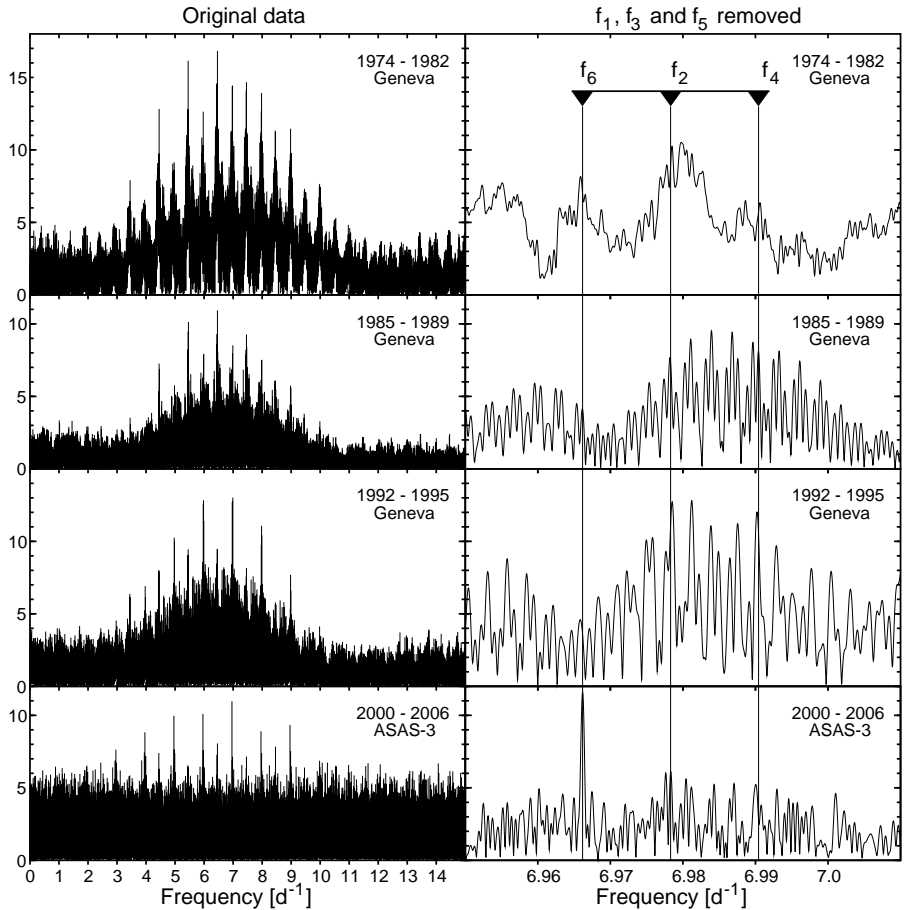}
\caption{{\it Left:} Fourier periodograms of the $V$-filter data of V836~Cen
calculated separately for four subsets defined in the text. The ordinate is
the amplitude in mmag. {\it Right:} Periodograms of the same subsets prewhitened
with $f_1$, $f_3$, and $f_5$. Only a narrow frequency range around 6.98~d$^{-1}$
is shown.  This is the region where the remaining three modes, with equally split
frequencies, occur. They are labelled as $f_6$, $f_2$, $f_4$ and indicated with
thin vertical lines.  Note the differences in the frequency pattern, which we
interpret as evidence for amplitude changes.}
\label{v836cen-per}
\end{figure*}

\begin{figure*}
\centering
\includegraphics[width=15.8cm]{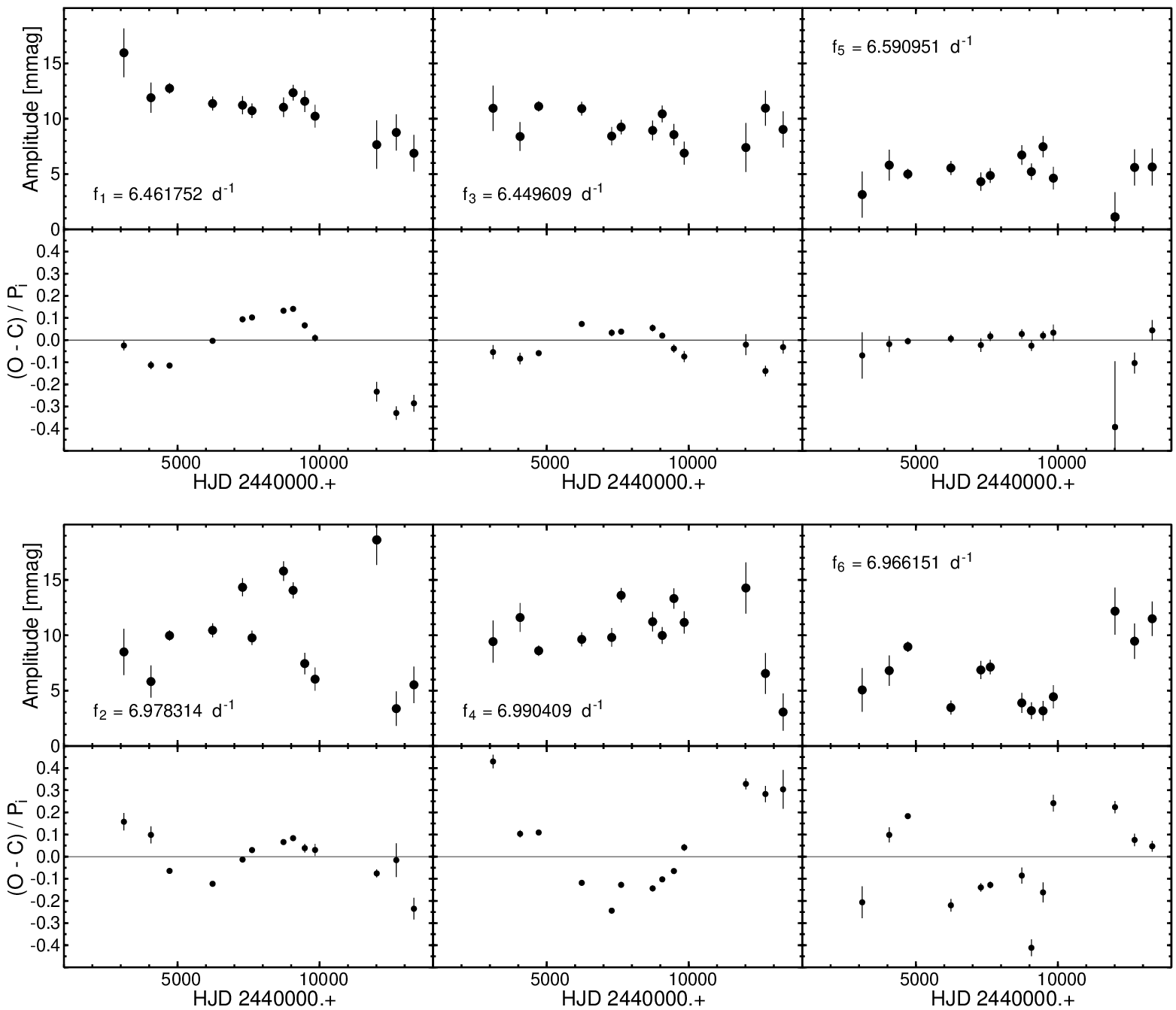}
\caption{$V$-filter semi-amplitudes and O$-$C diagrams for six modes of V836~Cen.
The ordinate range in the O$-$C diagrams equals the length of the
corresponding period.}
\label{v836cen-aoc}
\end{figure*}

In view of the suspected amplitude changes in V836~Cen, we decided to re-analyse
the Geneva data and check the stability of the amplitudes of the modes. Along
with the ASAS-3 photometry made in the years 2000--2006,  the $V$-filter data
for this star already cover about 30 years and are reasonably well distributed
over this interval.  First, we split the $V$-filter Geneva data into three
subsets covering the intervals: 1974--1982,  1985--1989, and 1992--1995, the
fourth set consisting of the ASAS-3 photometry. Each set was then analysed
independently.  We fitted all six modes with frequencies given by A04. In
addition, we allowed {\it linear} amplitude changes within each subset. The
amplitudes, the rates of amplitude change, phases and frequencies were
improved by means of a non-linear least-squares fit.  At this step, we also
removed some year-to-year trends present in the Geneva data as they produced
high signal at low frequencies in the periodograms.

The analysis showed convincingly that (i) there are large differences in the
amplitudes of modes for different subsets, as can be seen in
Fig.~\ref{v836cen-per}, (ii) the rates of amplitude change are significantly
different from zero for some mode/subset combinations. Note that the differences
in frequency patterns for two different datasets of V836 Cen were already
indicated by \cite{heyn92} when he compared his results with those of
\citet{waru83}.

\onltab{5}{%
\begin{table*}
\caption{Mean amplitude ratios for the six modes of V836~Cen.  Numbers in
parentheses denote r.m.s.~errors of the preceding numbers with the preceding
zeroes omitted.}
\label{v836cen-rat}
\begin{tabular}{ccccccccll}
\hline \hline
Mode & Frequency [d$^{-1}$]& $A_{\rm B1}/A_{\rm U}$ &  $A_{\rm B}/A_{\rm U}$ & 
$A_{\rm B2}/A_{\rm U}$ & $A_{\rm V1}/A_{\rm U}$ & $A_{\rm V}/A_{\rm U}$ & $A_{\rm G}/A_{\rm U}$ \\
\hline
$f_1$ & 6.461752(18) & 0.838(32) & 0.829(31) & 0.783(26) & 0.754(27) & 0.805(30) & 0.766(31)\\
$f_2$ & 6.978314(12) & 0.717(20) & 0.716(18) & 0.703(19) & 0.703(37) & 0.664(32) & 0.651(26)\\
$f_3$ & 6.449609(07) & 0.871(26) & 0.852(28) & 0.865(18) & 0.776(20) & 0.775(25) & 0.769(39)\\
$f_4$ & 6.990409(26) & 0.746(27) & 0.774(33) & 0.692(26) & 0.676(55) & 0.699(24) & 0.686(36)\\
$f_5$ & 6.590951(13) & 0.633(24) & 0.590(32) & 0.532(39) & 0.548(32) & 0.518(21) & 0.501(34)\\
$f_6$ & 6.966151(18) & 0.646(41) & 0.632(65) & 0.666(31) & 0.585(42) & 0.631(43) & 0.591(45)\\
\hline
\end{tabular}
\end{table*}
}

The question which arises now is the time scale of the amplitude changes. The
periodogram of the residuals from the solution mentioned above showed
significant peaks close to frequencies that were removed, leading us to suspect
that the time scale of the amplitude changes might be shorter than the length
of a subset, i.e., 3--8 years. The other reason for the presence of these
peaks might be period changes. In order to trace the changes of amplitudes on
a shorter time scale, we proceeded in the following way.  We used the solutions
mentioned above, i.e., solutions that allowed linear amplitude changes and were
calculated for each subset independently, to remove the contribution from all
modes but $f_1$. The residuals were then split into 13 shorter subsets, in most
cases covering a single season. For each such subset, a sinusoid with frequency
equal to $f_1$ was fitted in order to derive the amplitude and the time of
maximum light. A similar procedure was then applied to all six modes. The
resulting semi-amplitudes are shown in Fig.~\ref{v836cen-aoc}. We stress that
the procedure we used is not equivalent to fitting a six-mode model to the
seasonal data.  This would lead to the well-known problems with resolution and
aliasing.  Instead, it takes advantage of the fact that the frequency contents
is known from the best data and that--when we are left with only one mode not
subtracted--its amplitudes and phases can be traced in a yearly time scale. In
addition, it can be believed that the other modes are subtracted quite correctly
because they were all included in the model used for subtraction. This model,
in turn, was fitted to the dataset which had the time-span long enough to avoid
problems with frequency resolution.

It can be clearly seen from Fig.~\ref{v836cen-aoc} that the amplitude is
roughly constant only for $f_3$ and $f_5$. For $f_1$, it is decreasing steadily,
while for the triplet components, $f_2$, $f_4$, and $f_6$, it seems to change
less regularly, and large differences in amplitude can be seen from season to
season.

We performed another cross-check of the amplitude changes using two subsets of
the Geneva data, 1988--1989 and 1992--1995, and consecutively extracted modes
that appeared in the periodograms.  As expected, for each subset the modes were
recovered in the order consistent with the amplitudes shown in
Fig.~\ref{v836cen-aoc}. In addition, a double peak was found close to $f_2$ in
the periodogram of the 1992--1995 data.  This is simply a consequence of a
large drop in amplitude between 1992/3 and 1994/5 (Fig.~\ref{v836cen-aoc}).
The amplitude change also explains the occurrence of a double peak near $f_2$
in the analysis of the ASAS-3 data (see Fig.~\ref{v836cen-per}) as the
amplitude of $f_2$ has dropped significantly in the time interval covered by
these observations.

\begin{figure}
\centering
\includegraphics[]{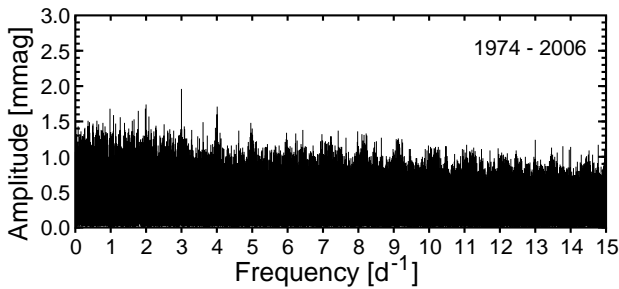}
\caption{Fourier periodogram of the residuals from the final solution of 
V836~Cen data in the $V$ filter.}
\label{v836cen-resid}
\end{figure}

Having derived the amplitudes and phases, we were left with the residuals that
were used to check the correctness of the fit. The Fourier periodogram of these
residuals is shown in Fig.~\ref{v836cen-resid}. The highest peak occurs at
frequency 3~d$^{-1}$ and is likely to be due to the residual extinction
unaccounted for in the Geneva data. No other significant peaks appear above
the 1.5~mmag level, even in the region where the frequencies of the six modes
were found (6.5--7 d$^{-1}$). We therefore conclude that the changes of
amplitudes and phases of the six modes shown in Fig.~\ref{v836cen-aoc} fully
describe the photometric variability of V836~Cen.  The occurrence of residual
power and additional low-amplitude modes ($f_7$ and $f_8$) in the 6--7~d$^{-1}$
range, mentioned by A04, were simply a consequence of neglecting the
possibility of amplitude changes.

Since the amplitudes of the pulsation modes in V836 Cen were used for mode
identification, it is reasonable to check if the changes of amplitudes and
phases in other filters follow the pattern found in the $V$ filter.  We
therefore made a similar analysis of the Geneva data in the remaining six
filters. The results confirm the character of the changes seen in the $V$ filter.
Except for the six modes found by A04, no others were detected. The detection
threshold in the 5--10~d$^{-1}$ range, defined as the mean amplitude in the
Fourier spectrum of residuals multiplied by 4, was equal to 2.2~mmag in $U$,
1.7~mmag in $B$, $B_1$, $B_2$ and $V_1$, 1.4~mmag in $V$, and 1.8~mmag in $G$.
The semi-amplitudes and times of maximum light are presented in Table 
\ref{v836cen-at}, and are available as online material.

\onltab{6}{
\begin{table*}
\caption{Semi-amplitudes and the times of maximum light derived from
sine-curve fits to 13 short datasets of $V$-filter observations of V836~Cen.
Repeatable integral parts of the epochs are replaced by `---' for $f_2$ to
$f_6$. The r.m.s.~errors are given underneath the values. See text for details.}
\label{v836cen-at}
{\small
\begin{tabular}{crrrrrrrrrrrrr}
\hline\hline
 & \multicolumn{6}{c}{Semi-amplitudes [mmag]}&& \multicolumn{6}{c}{$T_{\rm max}^i -$ HJD 2440000.0}\\ 
\cline{2-7}\cline{9-14}
Subset &\multicolumn{1}{c}{$f_1$}&\multicolumn{1}{c}{$f_2$}&\multicolumn{1}{c}{$f_3$}&\multicolumn{1}{c}{$f_4$}&\multicolumn{1}{c}{$f_5$}&
\multicolumn{1}{c}{$f_6$}&&\multicolumn{1}{c}{$f_1$}&\multicolumn{1}{c}{$f_2$}&\multicolumn{1}{c}{$f_3$}&\multicolumn{1}{c}{$f_4$}&
\multicolumn{1}{c}{$f_5$}&\multicolumn{1}{c}{$f_6$}\\
\hline
 1&16.0& 8.5&10.9& 9.4& 3.1& 5.1&&3110.6581&---.6591&---.6258&---.7113&---.7270&---.7106\\
&2.2&2.1&2.1&1.9&2.1&2.0&&0.0034&0.0056&0.0049&0.0044&0.0159&0.0103\\
 2&11.9& 5.8& 8.4&11.6& 5.8& 6.8&&4062.3987&---.3132&---.3071&---.3975&---.3427&---.3559\\
&1.4&1.5&1.3&1.3&1.4&1.4&&0.0028&0.0055&0.0040&0.0024&0.0055&0.0049\\
 3&12.7&10.0&11.1& 8.6& 5.0& 9.0&&4717.3292&---.3192&---.2345&---.2959&---.3337&---.2491\\
&0.5&0.5&0.5&0.5&0.5&0.5&&0.0009&0.0011&0.0010&0.0012&0.0022&0.0012\\
 4&11.4&10.5&10.9& 9.6& 5.6& 3.5&&6233.3441&---.4365&---.4713&---.3412&---.3536&---.3801\\
&0.6&0.6&0.6&0.6&0.6&0.6&&0.0014&0.0014&0.0014&0.0015&0.0027&0.0041\\
 5&11.2&14.3& 8.4& 9.8& 4.3& 6.9&&7288.4910&---.4349&---.4131&---.4836&---.4324&---.3502\\
&0.8&0.9&0.8&0.8&0.8&0.8&&0.0018&0.0013&0.0024&0.0019&0.0046&0.0028\\
 6&10.7& 9.8& 9.2&13.6& 4.9& 7.1&&7621.6837&---.7594&---.7675&---.6712&---.7743&---.6779\\
&0.7&0.7&0.7&0.7&0.7&0.7&&0.0015&0.0015&0.0017&0.0011&0.0032&0.0021\\
 7&11.0&15.8& 8.9&11.2& 6.7& 3.9&&8730.3661&---.3419&---.3647&---.3312&---.2656&---.3307\\
&0.9&0.9&0.9&0.9&0.9&0.9&&0.0019&0.0013&0.0024&0.0018&0.0032&0.0052\\
 8&12.3&14.1&10.4&10.0& 5.2& 3.2&&9068.8205&---.6783&---.6745&---.8009&---.7520&---.7778\\
&0.7&0.7&0.8&0.8&0.7&0.8&&0.0015&0.0012&0.0018&0.0017&0.0036&0.0054\\
 9&11.6& 7.4& 8.6&13.3&7.5& 3.2&&9478.6050&---.6563&---.6129&---.6537&---.7150&---.6527\\
&1.0&1.0&1.0&0.9&1.0&0.9&&0.0020&0.0028&0.0028&0.0016&0.0030&0.0065\\
10&10.2& 6.0& 6.9&11.2& 4.6& 4.4&&9842.4294&---.3533&---.3504&---.3102&---.3975&---.3260\\
&1.0&1.1&1.0&1.0&1.0&1.1&&0.0025&0.0039&0.0038&0.0022&0.0056&0.0054\\
11& 7.7&18.6&7.4&14.3& 1.1&12.2&&12013.9393&---4.0663&---3.9638&---4.0419&---3.9470&---3.9675\\
&2.2&2.3&2.2&2.3&2.2&2.1&&0.0068&0.0029&0.0073&0.0035&0.0451&0.0040\\
12& 8.8& 3.4&11.0& 6.5& 5.6& 9.5&&12702.7469&---.7800&---.8245&---.8365&---.8142&---.8489\\
&1.6&1.6&1.6&1.9&1.6&1.6&&0.0047&0.0109&0.0037&0.0052&0.0071&0.0040\\
13& 6.9& 5.5&9.0&3.1& 5.6&11.5&&13323.9475&---.9588&---.9643&---.9765&---.9914&---.9913\\
&1.7&1.7&1.6&1.7&1.7&1.6&&0.0059&0.0070&0.0045&0.0125&0.0070&0.0035\\
\hline
\end{tabular}

}
\end{table*}}

The parameters, particularly the amplitudes in different filters derived by A04,
were later used for seismic modelling of V836~Cen \citep{dupr04}. Since all
Geneva data were made simultaneously in all seven filters, one may expect that
the fact that their amplitudes are changing did not affect the amplitude ratios.
This is indeed the case.  For completeness, in Table \ref{v836cen-rat} we
provide the amplitude ratios calculated as weighted means of the ten ratios
obtained from the subsets covered by the Geneva data.

\subsection{\object{BW Cru} = \object{HIP\,62949} = NGC\,4755-F}
The star, a member of the open cluster NGC\,4755, was noted to have variable
radial velocity by \citet{hern60}. Its photometric variability was discovered
by \citet{jaka78} who found a period of 0.203~d, corresponding to frequency of
about 4.9~d$^{-1}$. The star was later studied by \citet{shob84} and
\citet{koen93} who confirmed its variability. \citet{koen93} found two
periodic terms with frequencies 4.89 and 6.16~d$^{-1}$. The latter frequency
was not found in the data analysed by \citet{bako94} and \citet{stan02}. The
accurate photometry of \citet{stan02} resulted in the confirmation of the
known mode with frequency of 4.88~d$^{-1}$. \citet{stan02} also discovered
two new modes with frequencies 4.54 and 5.27~d$^{-1}$.

In the ASAS-3 data, we detect only the main mode with frequency 4.88458 $\pm$
0.00004~d$^{-1}$ and the $V$-filter semi-amplitude equal to 10.3 $\pm$ 1.2~mmag
(Fig.~\ref{om2}).  The other modes are not detected because our detection
threshold amounts to about 6~mmag. It is interesting to note that the amplitude
of the main mode, despite probable contamination by nearby stars, is larger
than the values derived previously: 7~mmag \citep{koen93}, 6.4 $\pm$ 0.5~mmag
\citep{bako94}, 6.4 $\pm$ 0.8 \citep{balo97}, but agrees quite well with the
value of 9.6 $\pm$ 1.1~mmag, derived by \citet{stan02}.  We conclude that the
main mode in BW Cru undergoes a long-term increase of amplitude.

\subsection{\object{V348 Nor} = \object{HD\,147985} = \object{HIP\,80563}}
The star was found to be $\beta$~Cephei-type variable by \citet{wacu85}. These
authors found three periodicities with frequencies 7.5579, 6.8999, and
6.3834~d$^{-1}$ in their 1983 observations, carried out in the seven-band
Geneva system. The $V$-filter semi-amplitudes of these three modes were equal
to 23, 12.5 and 7 mmag, respectively.

In the ASAS-3 data, we detect only two modes (Fig.~\ref{om3}). The first
detected mode is the mode at 6.90033~d$^{-1}$ which has semi-amplitude equal
to 10.8 $\pm$ 0.8 mmag, in reasonable agreement with 12.5~mmag in the Geneva
data.  The other, with frequency of 7.55795~d$^{-1}$, has an amplitude of only
5.4 $\pm$ 0.8\,mmag, i.e., about four times smaller than in the discovery data.
We do not detect the third mode found by \citet{wacu85} despite the fact that
the detection threshold in the ASAS-3 photometry amounts to about 4\,mmag, well
below 7\,mmag, the semi-amplitude of the third mode in the 1983 data.
Apparently, the amplitudes of the two modes in V348~Nor, in particular, the
main mode, declined between 1983 and the epochs of the ASAS-3 observations
(2001--2006).  This is confirmed by the Hipparcos data, where only the main
mode with frequency 7.55778~d$^{-1}$ is detected. Its $H_{\rm p}$ semi-amplitude
amounts to about 16~mmag.  The other modes are not detected, but the detection
threshold is rather high in the Hipparcos data, amounting to about 9~mmag.

\section{Discussion}
It is not well known how common amplitude and period changes are among
$\beta$~Cephei stars.  The main reason for this is the lack of homogeneous
observations covering long time intervals. From the analysis of observations
for stars with the longest observational records it seems, however, that,
if detectable, changes of periods and amplitudes of modes excited in
$\beta$~Cephei stars have a typical time scale longer than a few months. In
many cases it is much longer than a decade as in the well-documented case
of 16 (EN) Lac \citep{jepi96, jepi99}.

\subsection{Changes of periods}
The evidence for long-term period variations among $\beta$~Cephei stars was
summarised a few years ago by \citet{jepi98} and \citet{jerz99}. One of the
main reasons for the long-term monitoring of periods in $\beta$~Cephei stars
is the hope that evolutionary changes in these stars will be detected and
confronted with the predictions of the theory.  According to the theory
\citep{egpe73, leai74}, in the core-hydrogen burning stage of evolution, where
most $\beta$~Cephei stars are believed to occur, the rate of period change,
$\dot{P}$, should be smaller than 0.3~seconds per century and positive.
Although this value is small, it is detectable provided that observations cover
at least a decade. $\dot{P} >$ 0 arises as a consequence of increase of stellar
radius during the evolution on the main sequence.

Stars with small positive $\dot{P}$, consistent with theoretical predictions
for core-hydrogen burning phase, are indeed observed \citep{jerz99}, although
there are cases (BW Vul, $\sigma$ Sco) where $\dot{P}$ is much larger than
0.3~s\,cen$^{-1}$ which was used as an argument in favour of the hypothesis
that they had already evolved off the main sequence \citep{pigu92,pigu93}.
In this group, there was also $\delta$~Cet. For this star, $\dot{P}$ was equal
to 0.47 $\pm$ 0.09~s\,cen$^{-1}$ \citep{jerz88}, slightly too large for the
main-sequence evolution \citep[see Fig.~1 of ][]{jerz99}. In view of the new
observations, however, it is clear that for this star the change of period can
no longer be interpreted in terms of a constant $\dot{P}$ \citep{jerz07}.

There are, however, $\beta$~Cephei stars in which the period behaviour was much
more complex. \citet{jerz99} listed three such stars: $\beta$~CMa, 12 (DD) Lac,
and 16 (EN) Lac. Indeed, the evolutionary changes of period may contribute to
the observed period changes in these stars only partly because the changes are
different for different modes. Moreover, for some modes the period was found
subsequently to increase and decrease. The pattern of period changes for modes
excited in V836~Cen (Fig.~\ref{v836cen-aoc}) resembles that observed in the
three above-mentioned stars, especially 12 Lac \citep{pigu94}. Thus, V836\,Cen
may be regarded as the fourth member of this group of $\beta$~Cephei stars
with complex period behaviour. It will be shown in Paper II that one of the
new $\beta$~Cephei stars, HD\,168050, also exhibits a complex period behaviour.
All five stars from this group are multiperiodic. It seems therefore that when
many modes are excited, some kind of mode interaction causes period and, as we
shall comment below, amplitude changes.

In this context it is worth noting that in the study of period variations we
assume that pulsations are coherent. In a general case, this might not be true.
If random phase shifts are generated in a star, e.g., by the presence of other
mode(s), period variations similar to those seen in 12 Lac and V836\,Cen can be
observed. For such stars we can expect to detect evolutionary period changes
only on a time scale which would average random changes. For $\beta$~Cephei
stars this probably means centuries.

Period variations in $\beta$~Cephei stars might be additionally complicated
by the presence of a star in a wide binary system. In such a case, additional
contribution to the apparent period changes comes from the light-time effect.
This is indeed observed in $\beta$~Cep \citep{pibo92}, $\sigma$~Sco
\citep{pigu92} and possibly BW Vul \citep{pigu93}. As KK Vel is known to have
a close visual companion at separation of about 0$\farcs$3, the large changes
of period seen in Fig.~\ref{kkvel-pc}, apparently in both directions, can be
mostly due to the light-time effect.

\subsection{Changes of amplitudes}
Changes of amplitudes were reported for only a handful of $\beta$~Cephei stars.
\citet{jerz99} listed five stars with long-term amplitude variations:
$\beta$~CMa, 27 (EW)~CMa, V381~Car, $\alpha$~Vir A, and 16~Lac. Evidence for
amplitude change was also found for $\nu$~Cen \citep{aspa92,scte02} and for
one mode in 12 Lac \citep{pigu94}. As modes in V836\,Cen also show amplitude
changes (Fig.~\ref{v836cen-aoc}), we conclude that all stars with complex
period behaviour also show amplitude changes. The outstanding examples from
this group are the primary of Spica ($\alpha$~Vir) in which pulsation
amplitudes dropped to undetectable level or ceased altogether 
\citep{lomb78,ster86} and 16 Lac which was monitored photometrically many
times since the time of discovery. The changes of amplitudes and periods of
its four dominant modes are therefore known very well \citep{jepi96,jepi99}.
Apart from V836~Cen, we report amplitude changes for BW~Cru and V348~Nor.
Note that both BW~Cru and V348~Nor are not known as multiperiodic.

In this paper, combining ASAS-3 photometry with the ar\-chi\-val data, we found
evidence for period and amplitude changes in four known $\beta$~Cephei stars.
In addition, in Paper II, complex period changes in HD\,168050 were found using
only ASAS-3 data. This shows that period and amplitude changes might be quite
common in $\beta$~Cephei stars but, obviously, can be detected only if the data
cover sufficient time intervals, typically more than a decade. Fortunately,
ASAS is continuing and will be soon extended into the northern hemisphere.
In addition, other photometric surveys are expected to provide abundant
time-series data. In view of the fact that over 200 new $\beta$~Cephei stars
were detected in the ASAS-3 data, the sample of stars in which we will be able
to monitor long-term changes of amplitudes and periods will increase
considerably over the next decades.

\begin{acknowledgements}
The work was supported by the MNiI/MNiSzW grants No. 1 P03D 016 27 and N203
007 31/1328. We greatly acknowledge comments made by Prof.~M.\,Jerzykiewicz.
This research has made use of the SIMBAD database, operated at CDS, Strasbourg,
France.
\end{acknowledgements}

\Online

\begin{appendix}
\section{Notes on individual stars}
\subsection{Stars from the list of \citet{stha05}}
{\bf \object{V350 Pup} = \object{HD\,59864} = \object{HIP\,26500}.}
This star was discovered as a double-mode $\beta$~Cephei-type variable by
\citet{stje90}.  Unfortunately, due to aliasing problems, the frequencies of
the two modes they found, could not be indicated unambiguously.

The highest peak in the periodogram of the ASAS-3 data of V350~Pup slightly
exceeds the detection limit and occurs at frequency $f$ = 4.23945~d$^{-1}$
(Fig.~\ref{om1}). This frequency coincides perfectly with the fourth frequency
in the list of equally high peaks (4.240~d$^{-1}$) given by \citet{stje90}.
It is therefore very likely that this value represents the true pulsation
frequency or its daily alias.  The semi-amplitude in the ASAS-3 data amounts
to 3.9 $\pm$ 0.6~mmag which is less than the value of 7.1--7.7~mmag reported
by \citet{stje90}. The latter observations were, however, carried out in the
$B$ band, while the ASAS-3 data were obtained in the $V$ band, so that the
amplitudes are not directly comparable. No significant peaks were found in the
Hipparcos data for this star, but the detection threshold amounted to about
7~mmag.

{\bf \object{YZ Pyx} = \object{HD\,71913} = \object{HIP\,41586}.} The
variability of YZ Pyx was discovered in Hipparcos data \citep{wael98};
subsequently the star was studied by \citet{aert00}.  She found a single
frequency both in the Hipparcos and Geneva data, with the $V$ semi-amplitude
of 16 mmag. A single mode with a very similar amplitude was also found in the
ASAS-3 data (Fig.~\ref{om1}, Table \ref{bc-known-f}). All existing data
(Hipparcos, Geneva and ASAS-3) of YZ Pyx are well represented by a single mode
with a constant period of 0.2057818 $\pm$ 0.0000002~d.

{\bf \object{IL Vel} = \object{HD 80383}.} ASAS-3 photometry and the
variability record were already discussed by P05. However, since the time of
publication of that paper, some new data were obtained by ASAS-3.  Consequently,
we provide a new solution, essentially in agreement with that of P05.  The
frequency spectrum of IL Vel is dominated by two high-amplitude modes, the
only ones which were detected in the ASAS-3 data (Fig.~\ref{om1}). From the
analysis of the archival data we conclude that the amplitudes and periods of
these two main modes were stable during the last 30 years. The mean periods of
these modes are equal to 0.18315764 $\pm$ 0.00000003~d and 0.18645460 $\pm$
0.00000005~d. However, we confirm the suggestion made by \citet{hand03},
that more modes are present in this star. This can be judged primarily from
the analysis of the data of \citet{heha94}. Unfortunately, these data suffer
from a strong aliasing problem, so that the frequencies of the low-amplitude
modes cannot be given unambiguously. The star certainly deserves a follow-up
multi-site campaign, as it is potentially a very good target for asteroseismology.

{\bf \object{V433 Car} = \object{HD\,90288}.} This fast-rotating star was
discovered to be variable in light by \citet{lamp88}, and then studied by
\citet{heyn92} who found four periodic terms in his photometry.  Four modes
were confirmed by \citet{hand03}, although frequencies of some differed
by 1\,d$^{-1}$ from the values given by \citet{heyn92}.  \citet{hand03} also
found one combination frequency. In the ASAS-3 data, we find only the two
periodic terms with the largest amplitudes (Fig.~\ref{om1}). Their frequencies
differ slightly from those given by \citet{hand03}, but this may be a
consequence of the fact that we did not fit the modes with low amplitudes which
were below the detection threshold in the ASAS-3 data (4.6 mmag).
The amplitudes of the two main modes agree to within the errors with those
given by \citet{hand03}.

{\bf \object{Stars in NGC\,3293}.} Eleven $\beta$~Cephei stars are known in
this young open cluster \citep{balo77,baen83,heyn92,balo94,balo97}. All have
magnitudes in the range covered by ASAS-3 data. Unfortunately, because of the
limited spatial resolution of the ASAS-3 data (see Sect.~2), we detect
photometric variability due to pulsations in only two of them, V401\,Car and
V403\,Car.

\object{V401 Car} (NGC\,3293-10) was discovered by \citet{baen83} who reported
a period of 0.176~d. \citet{enge86} listed five modes, two of which were
confirmed later by \citet{heyn92}. \citet{balo94} also found two modes, but
frequency of only one agreed with that of \citet{heyn92}. Finally,
\citet{balo97} reported three frequencies. The frequencies given in these
papers do not always agree, even if aliasing is taken into account.  The
mode which is listed most frequently is the mode with the highest amplitude,
equal to about 10 mmag in $V$, having a frequency of 5.92~d$^{-1}$. We detect
it in the ASAS-3 data, although at the detection limit (S/N = 4.0, see
Fig.~\ref{om1}) and with amplitude of only 6 mmag, apparently lowered by
contamination.

\object{V403 Car} (NGC\,3293-16) was discovered by \citet{baen83} as a star
pulsating with a period of 0.249~d. \citet{enge86} found the same dominant mode
with frequency of 3.99~d$^{-1}$ and a secondary one at $f$ = 4.92~d$^{-1}$.
\citet{heyn92} re-analysed data obtained by \citet{cuyp85} and found also two
modes, with frequencies 3.9929 and 3.9147~d$^{-1}$.  Finally, \citet{balo94}
and \citet{balo97} found a single mode with a frequency of 3.996 and
3.990~d$^{-1}$, respectively. The main mode has the $V$-filter semi-amplitude
of about 25~mmag.  We detect it in the ASAS-3 data (Fig.~\ref{om2}), although
with smaller amplitude evidently due to contamination by nearby stars. As the
yearly aliases are rather small in the ASAS-3 data, we conclude that the value
of 3.9929~d$^{-1}$ reported by \citet{heyn92}, is a yearly alias of the true
frequency. The secondary mode is possibly real as it was detected in two
independent data sets; its frequency found by \citet{heyn92}, is a daily alias
of that given by \citet{enge86}. It could not be detected in the ASAS-3 data
due to its small amplitude.

\begin{figure}
\centering
\includegraphics[width=8cm]{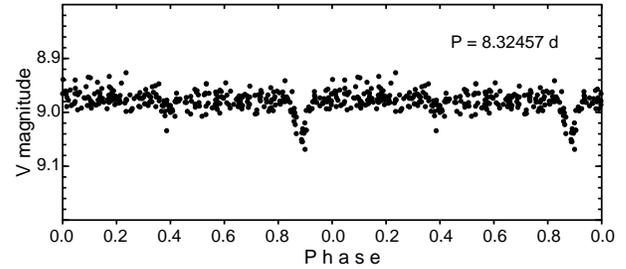}
\caption{ASAS-3 data of V381~Car phased with the orbital period of 8.32457~d
given by \citet{frey05}.}
\label{v381car}
\end{figure}

\object{V381 Car} (NGC\,3293-5) = \object{HD\,92024} is an 8.32-day detached
eclipsing system with a $\beta$~Cephei-type primary. It was discovered by
\citet{enge86} and then studied by \citet{enba86}, \citet{jest92} and recently,
by \citet{frey05}. We do not detect pulsations in this star. However, the
eclipses, with lowered depths, again due to contamination, can be seen in the
phased ASAS-3 light curve (Fig.~\ref{v381car}).

{\bf \object{KZ Mus} = \object{HD\,109885} = \object{HIP\,61751}.}  The star
was noted as variable by \citet{tobi94}. The variability was confirmed in the
Hipparcos data \citep{wael98,aert00}.  The most recent study of the star was
that of \citet{hand03} who found four independent modes in the multicolour
photometry of KZ Mus.  We find three modes in the ASAS-3 data (Fig.~\ref{om2}).
Their amplitudes and frequencies agree very well with those of \citet{hand03},
see Table \ref{bc-known-f}. We did not detect the fourth mode; the frequency
of the highest peak in the periodogram of residuals, however, equal to
5.7103~d$^{-1}$, is very close to the frequency of the fourth mode found by
\citet{hand03}.

{\bf \object{Stars in NGC 4755}.} This is another young open cluster in the
southern hemisphere known to contain a large number, namely ten,
$\beta$~Cephei-type stars \citep{jaka78,shob84,koen93,bako94,stan02}. We
find pulsations in only one $\beta$~Cephei star in this cluster, namely 
BW\,Cru, already discussed in Sect.~5. In addition, for CV Cru = NGC\,4755-I,
the known mode with $f$ = 5.585~d$^{-1}$ is marginally detected with S/N = 3.8.
For the remaining stars, the ASAS-3 photometry either does not exist or has
large photometric errors.

{\bf V\object{856 Cen} = \object{HD\,112481} = \object{HIP\,63250}.}
The variability of V856\,Cen was discovered by \citet{wahe89}. They found two
periodic terms. The first had a frequency of 3.9287 d$^{-1}$, but the frequency
of the other, 3.851~d$^{-1}$, was uncertain. The star was subsequently studied
by \citet{heyn92}, who confirmed both modes and derived their frequencies,
3.92870 and 3.85181~d$^{-1}$.  We detected only the main mode (Fig.~\ref{om2})
with frequency 3.92867 $\pm$ 0.00002~d$^{-1}$, in very good agreement with
previous determinations.  The main mode is also easily detected in the
Hipparcos data with a frequency of 3.92860 $\pm$ 0.00005~d$^{-1}$.

{\bf \object{V349 Nor} = \object{HD\,145794}}. The variability of HD\,145794
was discovered by \citet{wahe89}. Two periodic terms were found. The first had
frequency of 6.2507 d$^{-1}$; for the other, due to aliasing, there were two
possibilities: 4.2101 or 5.2129\,d$^{-1}$. We find the main mode with frequency
6.25354~d$^{-1}$ (Fig.~\ref{om2}), corresponding to 1~year$^{-1}$ alias of the
frequency found by \citet{wahe89}. For the second mode, S/N = 4.3, barely
exceeding the limiting value. The frequency amounts to 5.21316\,d$^{-1}$,
in agreement with the second value given by the discoverers.

{\bf \object{Stars in NGC 6231}.} This is the third southern open cluster with
a large number (six) of known $\beta$~Cephei-type stars 
\citep{shob79,bash83,baen85,bala95,aren01}. As for NGC\,3293 and NGC\,4755,
the magnitudes of $\beta$~Cephei stars in this cluster are within the range
covered by the ASAS-3 observations. Unfortunately, we did not detect any
pulsations for these stars. For some of them the data were too sparse, for
the other, the photometric errors were too large.

{\bf \object{V1035 Sco} = \object{HD\,156327\,B} = \object{HIP\,84655}.}
\object{HD 156327} is a well-known WC7-type Wolf-Rayet star WR\,86. Due to the
presence of absorption lines in the spectrum, the star was suspected to have
a companion \citep{robe62,smit68}. The star was also noted as a visual binary
by \citet{jeff63}. The presence of a close companion was excluded because no
clear radial-velocity variations were found \citep{mass81}, but a visual\
component was confirmed by means of speckle interferometry \citep{hart93}, and
later with the HST \citep{niem98,lepi01}. The components, very similar in
brightness, are separated by about 0$\farcs$23. The secondary is a B0\,III
star \citep{lepi01}.

Photometric variability of HD\,156327 with a period of about 3 hours was
discovered by \citet{mond88} from a single-night observations in the Walraven
$WBLUV$ system.  Similar variability was observed by \citet{vgen90}, and the
variations were at first attributed to the WR-type primary. However,
\citet{vgen91} considered the possibility that the B0-type secondary is a
$\beta$~Cephei star and derived 0.1385 $\pm$ 0.0002~d ($f$ = 7.22~d$^{-1}$)
for the period of the photometric variations. Owing to the evident variability,
the star was named V1035~Sco \citep{kasa97}. A detailed photometric and
spectroscopic study was later performed by \citet{paar02}. They concluded
that the short-period photometric variability of V1035~Sco should be attributed
to the secondary component and can be described by two modes with frequencies
6.914 and 7.236~d$^{-1}$.

In the ASAS-3 data, which are quite numerous for this star, we detect
unambiguously three modes (Fig.~\ref{om3}, Table \ref{bc-known-f}). If the
frequency resolution of the observations made by \citet{paar02} is taken into
account, it can be stated that the first two modes are the same as those found
by \citet{paar02}.

A periodogram of Hipparcos data of V1035\,Sco shows only an increased power
in the region of the detected modes, but the photometry is too poor to detect
any mode.

{\bf \object{V831 Ara} = \object{HD\,156662}.} Photometric variability of
V831 Ara was discovered by \citet{wacu85} who found three modes in the Geneva
photometry of this star, with semi-amplitudes of 8, 6, and 5.5~mmag in $V$.
The frequencies were later re-calculated by R.~Peetermans, as reported by
\citet{heyn92}, and were found to be equal to 5.30269, 5.89065, and 5.92120~d$^{-1}$.

We also find three modes in the ASAS-3 photometry of this star (Fig.~\ref{om3}).
The first mode we detect has frequency $f_1$ = 6.30519~d$^{-1}$, a daily alias
of the frequency given by \citet{heyn92}. After prewhitening with $f_1$, the
two remaining modes were recovered as well, with frequencies equal to 5.88907
and 5.92138~d$^{-1}$, in agreement with \citet{heyn92}.  In fact, a daily alias
with frequency of about 3.89~d$^{-1}$ was slightly higher than that at
5.889~d$^{-1}$, but it was 2~d$^{-1}$ apart, so we adopted the value closest
to that of \citet{heyn92}.  Since both sources of data suffer from severe daily
aliasing, it is difficult to decide which of the alias frequencies for $f_1$ is
correct.

{\bf \object{V2371 Oph} = \object{HD\,157485} = \object{HIP\,85189}.} The star
was discovered in the Hipparcos data by \citet{aert00}. She found two modes
with frequencies equal to 4.521 and 4.464~d$^{-1}$, $V$ amplitudes of 24 and
14~mmag, identified as $\ell$ = 1 and a radial mode, respectively.  We also
detect two modes, with the same frequencies and amplitudes (Fig.~\ref{om3},
Table \ref{bc-known-f}).

{\bf \object{NSV\,24078} = \object{HD\,164340} = \object{HIP\,88352}.}
A possible variability of radial velocity of HD\,164340 was already noted by
\citet{kilk75}, and then confirmed by \citet{kihi75}.  The star's photometric
variability was found in the Hipparcos data. It was included in the Hipparcos
catalogue as an unsolved variable and designated as NSV\,24078. In the
Hipparcos photometry of this star, \citet{mopo05} found two periodic terms
with frequencies 6.53876 and 6.37776~d$^{-1}$.

The ASAS-3 photometry of NSV\,24078 was already analysed by P05 revealing the
same two modes and possibly a third one. With additional data from ASAS-3,
we provide a new solution here. The third mode is now confirmed
(Fig.~\ref{om3}); its frequency is a daily alias of that mentioned by P05.
In addition, the fourth mode, at frequency 7.74246~d$^{-1}$, was detected.

{\bf V4382 Sgr = HD\,165812 = HIP\,88884.} The star was discovered in the
Hipparcos photometry by \citet{wael98}. From Geneva photometry, \citet{aert00}
found two modes.  The frequency of the main mode, however, derived from the
Hipparcos data (5.9164~d$^{-1}$) differed from that obtained from the Geneva
photometry (5.686~d$^{-1}$).  With the ASAS-3 data, we confirm the two modes
found by \citet{aert00}.  They have practically the same amplitudes in the
ASAS-3 photometry as in the Geneva data. The frequency of the second mode is,
however, the daily alias of that found by \citet{aert00} and amounts to
5.58497~d$^{-1}$. In addition, we found a third mode (Fig.~\ref{om4},
Table \ref{bc-known-f}).

{\bf \object{V4159 Sgr} = \object{HD\,166540} = \object{HIP\,89164}.}
This B0.5\,IV-type star \citep{morg53} was discovered to be variable by
\citet{wael91}.  They found one dominant mode with a frequency of
4.2920~d$^{-1}$ and an indication for more modes with frequencies in the range
between 4.2 and 4.5~d$^{-1}$.  The secondary and tertiary modes reported by
\citet{wael91} had frequencies of 4.2303 and 4.3996~d$^{-1}$. We found five
modes in the ASAS-3 photometry of this star (Fig.~\ref{om4}). Although the
frequencies also fall in the range between 4.24 and 4.36~d$^{-1}$, none agrees
with those given by \citet{wael91}. The five modes form a close pair and a
triplet.  The pair is separated by 0.0144~d$^{-1}$, while the separations in
the triplet equal 0.0027 and 0.0052~d$^{-1}$.  The former value is equal to
1~yr$^{-1}$, the latter is roughly twice as long. If the modes are real, this
is unfavourable, although the yearly aliases in the spectral window of the
ASAS-3 data of V4159~Sgr are not very strong: about 60\% of the height of
the main peak. Nevertheless, the frequencies of the triplet should be treated
with caution.

Unfortunately, the Hipparcos data do not resolve the problem. They are too
few for detecting any mode unambiguously.  In the periodogram of the Hipparcos
data, only an excess of power can be seen in the range of frequencies where the
modes found with the ASAS-3 data occur.

{\bf \object{V1449 Aql} = \object{HD\,180642} = \object{HIP\,94793}.}
The ASAS-3 photometry was previously analysed by P05. With some new data
acquired since this publication of this paper, we present a new solution which
is in very good agreement with the previous one: a single, large-amplitude mode
is detected (Fig.~\ref{om4}, Table \ref{bc-known-f}).

{\bf \object{SY Equ} = \object{HD\,203664}.} This is another star found to be
variable by Hipparcos and then re-observed by \citet{aert00}. As for V1449~Aql,
its ASAS-3 photometry was previously analysed by P05. We provide a new solution for
this star, again in agreement with the previous one.  New Geneva photometry of
SY\,Equ was also recently obtained by \citet{aert06}. In addition to the known
high-amplitude mode, they detected two low-amplitude ones. The amplitudes are,
however, lower than the detection threshold in the ASAS-3 data (5.6~mmag), so
we were unable to confirm them.

{\bf \object{HN Aqr} = \object{PHL 346}.} This B1-type star is a very
interesting case because of its high Galactic latitude ($b$ = $-$58$\degr$)
and Population I abundances \citep{klbd86,ryan96}. It opened a discussion on
the possibility of star formation far outside the Galactic plane
\citep[e.g.][]{keen86, keen92}. For HN Aqr, it is still not certain if the
star was formed outside the Galactic plane \citep{hamb96,lynn02} or is a
runaway object \citep{rams01}.

HN~Aqr was found to be photometrically variable by \citet{waru88} who observed
the star in the Geneva system. They found a single mode with a period of
0.1522~d. This variability was confirmed by \citet{kivw90} in $BV$ photometry.
The variability was also detected in the ultraviolet flux and radial velocities
\citep{duft98}.

The analysis of the ASAS-3 data also reveals a single mode with the same
frequency and amplitude as found by previous investigators (Fig.~\ref{om4}).

\subsection{Stars found by \citet{pigu05}}
As mentioned in the Introduction, the published ASAS-3 catalogue already
contained a number of new $\beta$~Cephei stars. \citet{pigu05} verified the
results of the automatic classification finding that the catalogue contains
14 new $\beta$~Cephei stars. He also provided multi-frequency solutions for
these stars. Since the publication of that paper, data for the equatorial
stars were made available \citep{asas5}. In addition, new observations were
acquired for all stars.  We therefore updated the solutions for 14
$\beta$~Cephei stars found by P05. They are presented in Table
\ref{bc-known-ap}. We do not show updated periodograms because they are,
in general, very similar to those already presented by P05.

For seven stars, the solutions are practically the same as those provided
by P05. For the remaining seven stars, there are some differences: For
\object{HD\,133823}, the peak at about 0.5~d$^{-1}$ was suspected by P05
to be spurious. With the new data added, the peak stands out more clearly
in the periodogram, so we included it in the solution. In addition,
two other modes in its vicinity were detected.  The values of S/N for the
three low-frequency peaks amount to 6.4, 5.5, and 4.9, i.e., clearly above
the detection threshold. The frequencies of the three peaks are in the range
of 0.50--0.61~d$^{-1}$, suggesting either $g$-mode pulsations or variability
related to rotation.  In the former case, HD\,133823 would be an interesting
example of a hybrid $\beta$~Cephei/SPB type of variability.

For \object{CPD\,$-$50$\degr$9210} we find a third mode. It is interesting
to note that 2($f_3 - f_2$) $\approx$ ($f_2-f_1$).

A new mode was also found for \object{HDE\,328906}.  Due to aliasing, its
frequency is uncertain, however.  Although the peak at 8.264~d$^{-1}$ is the
highest, the daily aliases at 6.261 and 5.259~d$^{-1}$ are almost equally high.
We finally adopted the frequency of 5.259~d$^{-1}$ for $f_2$ because it is
closest to the frequency of the main mode. This needs to be verified with
future observations, however.

For \object{HD\,152077}, we decided not to include the $f_4$ mode listed by
P05 in the new solution because after prewhitening with the three modes and
the combination frequency, the aliasing problem becomes very severe. It is,
however, obvious, that at least two more low-amplitude modes are present in
this star.

For \object{HD\,155336}, P05 provided a solution with three modes. The new
solution also includes three modes, but two, $f_2$ and $f_3$, have frequencies
different from those given by P05. The difference is the result of a
complicated structure of the window function. This is because 60\% of the
ASAS-3 data for HD\,155336 were made during 6 observing nights spread over
20 days.  However, if these nights are omitted, the same three modes as given
in Table \ref{bc-known-ap} are detected. This strengthens the reliability of
the new solution.

There was also a problem with the correct identification of aliases for
\object{HD\,165582}, as discussed in detail by P05.  For\-tu\-na\-te\-ly, over
400 new datapoints were obtained for this star.  This allows us to give a new,
presumably correct, solution. The new $f_2$ is roughly equal to the old
$f_2$ minus 2~d$^{-1}$, whereas the new $f_3$ equals to the old $f_4$ plus
2~d$^{-1}$. The $f_2$ and $f_3$ modes are very close, the beat period
between them equals to about 250~days.

Finally, for \object{ALS\,5040}, the new solution substitutes a daily alias
for the old frequency $f_3$.

\subsection{Stars found by \citet{hand05i}}
As we mentioned in the Introduction, \citet{hand05i} identified five new
$\beta$~Cephei stars in the ASAS photometry, four in the $I$-filter ASAS-2 data
and one, \object{HD\,191531}, in the ASAS-3 data. The latter was not found by
P05 because the star was included in the published catalogue after his analysis.
Also, the four stars found by \citet{hand05i} in the ASAS-2 data were not listed
as variable in the published ASAS-3 catalogue of variable stars. We analyse
their ASAS-3 photometry in this paper. The stars are included in Table
\ref{bc-known}, the parameters of the sine-curve fits to the ASAS-3 photometry
are given in Table \ref{bc-known-gh}, and their Fourier periodograms are shown
in Fig.~\ref{om-gh}.

For \object{HD\,100495} = \object{ALS\,2386} we find the same mode as
\citet{hand05i} in the ASAS-2 data, but, in addition, two low-amplitude modes.
The frequency $f_3$ is probably the daily alias of the $f_2$ mode suspected
by \citet{hand05i}. For \object{CPD $-$61$\degr$3314} = \object{ALS\,2714}
we detect the same four modes and a combination frequency as \citet{hand05i}
but in addition, another combination mode, $f_2$ + $f_3$. It is interesting
that all modes have relatively large amplitudes for a $\beta$~Cephei star.
For each of the remaining three stars, \object{ALS\,2798}, \object{ALS\,2877},
and \object{HD\,191531} = \object{HIP\,99327}, we find one mode less than
\citet{hand05i}.

\end{appendix}

\begin{thebibliography}{}
\bibitem[Aerts(2000)]{aert00} Aerts, C. 2000, \aap, 361, 245
\bibitem[Aerts et al.(2004a)]{aert04a} Aerts, C., De Cat, P., Handler, G., et al. 2004a, \mnras, 347, 463
\bibitem[Aerts et al.(2004b)]{aert04} Aerts, C., Waelkens, C., Daszy\'nska-Daszkiewicz, J., et al. 2004b, \aap, 415, 241 (A04)
\bibitem[Aerts et al.(2006)]{aert06} Aerts, C., De Cat, P., De Ridder, J., et al. 2006, \aap, 449, 305
\bibitem[Arentoft et al.(2001)]{aren01} Arentoft, T., Sterken, C., Knudsen, M.R., et al. 2001, \aap, 380, 599
\bibitem[Ashoka \& Padmini(1992)]{aspa92} Ashoka, B.N., \& Padmini, V.N. 1992, \apss, 192, 79
\bibitem[Balona(1977)]{balo77} Balona, L.A. 1977, \memras, 84, 101
\bibitem[Balona(1994)]{balo94} Balona, L.A. 1994, \mnras, 267, 1060
\bibitem[Balona \& En\-gel\-brecht(1983)]{baen83} Balona, L.A., \& Engelbrecht, C. 1983, \mnras, 202, 293
\bibitem[Balona \& En\-gel\-brecht(1985)]{baen85} Balona, L.A., \& Engelbrecht, C. 1985, \mnras, 212, 889
\bibitem[Balona \& Koen(1994)]{bako94} Balona, L.A., \& Koen, C. 1994, \mnras, 267, 1071
\bibitem[Balona \& Laney(1995)]{bala95} Balona, L.A., \& Laney, C.D. 1995, \mnras, 276, 627
\bibitem[Balona \& Shobbrook(1983)]{bash83} Balona, L.A., \& Shobbrook, R.R. 1983, \mnras, 205, 309
\bibitem[Balona et al.(1997)Balona, Dziembowski \& Pamyatnykh]{balo97} Balona, L.A., Dziembowski, W.A., \& Pamyatnykh, A. 1997, \mnras, 289, 25
\bibitem[Cousins(1982)]{cous82} Cousins, A.W.J. 1982, IBVS, 2158
\bibitem[Cou\-sins \& Stoy(1962)]{cost62} Cousins, A.W.J., \& Stoy, R.H. 1962, Royal Obs. Bull., No. 49
\bibitem[Cuypers(1985)]{cuyp85} Cuypers, J. 1985. Ph.D.~thesis, Katholieke Univ.~Leuven
\bibitem[Dufton et al.(1998)]{duft98} Dufton, P.L., Keenan, F.P., Kilkenny, D., et al. 1998, \mnras, 297, 565
\bibitem[Dupret et al.(2004)]{dupr04} Dupret, M.-A., Thoul, A., Scuflaire, R., et al. 2004, \aap, 415, 251
\bibitem[Dziembowski \& Pamyatnykh(1993)]{dzpa93} Dziembowski, W.A., \& Pamyatnykh, A.A. 1993, \mnras, 262, 204
\bibitem[Eggleton \& Percy(1973)]{egpe73} Eggleton, P.P., \& Percy, J.R. 1973, \mnras, 161, 421
\bibitem[Engelbrecht(1986)]{enge86} Engelbrecht, C. 1986, \mnras, 223, 189
\bibitem[Engelbrecht \& Balona(1986)]{enba86} Engelbrecht, C., \& Balona, L.A. 1986, \mnras, 219, 449
\bibitem[Evans et al.(2005)]{evan05} Evans, C.J., Smartt, S.J., Lee, J.-K., et al. 2005, \aap, 437, 467
\bibitem[Eyer \& Blake(2005)]{eybl05} Eyer, L., \& Blake, C. 2005, \mnras, 358, 30
\bibitem[FitzGerald(1987)]{fitz87} FitzGerald, M.P. 1987, \mnras, 229, 227
\bibitem[Freyhammer et al.(2005)]{frey05} Freyhammer, L.M., Hensberge, H., Sterken, C., et al. 2005, \aap, 429, 631
\bibitem[Garrison et al.(1977)Garrison, Hiltner \& Schild]{garr77} Garrison, R.F., Hiltner, W.A., \& Schild, R.E. 1977, \apjs, 35, 111
\bibitem[Hambly et al.(1996)]{hamb96} Hambly, N.C., Wood, K.D., Keenan, F.P., et al. 1996, \aap, 306, 119
\bibitem[Handler(2005)]{hand05i} Handler, G. 2005, IBVS, 5667
\bibitem[Handler et al.(2003)]{hand03} Handler, G., Shobbrook, R.R., Vuthela, F.F., et al. 2003, \mnras, 341, 1005
\bibitem[Hartkopf et al.(1993)]{hart93} Hartkopf, W.I., Mason, B.D., Barry, D.J., McAlister, H.A., \& Bagnuolo, W.G. 1993, \aj, 106, 352
\bibitem[Hern\'andez(1960)]{hern60} Hern\'andez, C. 1960, \pasp, 72, 416
\bibitem[Heyn\-de\-rickx(1992)]{heyn92} Heynderickx, D. 1992, \aaps, 96, 207
\bibitem[Heyn\-de\-rickx \& Haug(1994)]{heha94} Heynderickx, D., \& Haug, U. 1994, \aaps, 106, 79
\bibitem[Hill(1970)]{hill70} Hill, P.W. 1970, \mnras, 150, 23
\bibitem[Hill et al.(1974)Hill, Kilkenny \& van Breda]{hill74} Hill, P.W., Kilkenny, D., \& van Breda, I.G. 1974, \mnras, 168, 451
\bibitem[Houk(1978)]{houk2} Houk, N. 1978, Michigan Spectral Survey, Ann Arbor, Dep. Astron., Univ. Michigan, 2
\bibitem[Houk(1982)]{houk3} Houk, N. 1982, Michigan Spectral Survey, Ann Arbor, Dep. Astron., Univ. Michigan, 3
\bibitem[Houk \& Cowley(1975)]{houk1} Houk, N., \& Cowley, A.P. 1975, Michigan Spectral Survey, Ann Arbor, Dep. Astron., Univ. Michigan, 1
\bibitem[Houk \& Smith-Moore(1988)]{houk4} Houk, N., \& Smith-Moore, M. 1988, Michigan Spectral Survey, Ann Arbor, Dep. Astron., Univ. Michigan, 4
\bibitem[Jakate(1978)]{jaka78} Jakate, S.M. 1978, \aj, 83, 1197
\bibitem[Jeffers et al.(1963)Jeffers, van den Bos \& Greeby]{jeff63} Jeffers, H.M., van den Bos, W.H., \& Greeby, F.M. 1963, Index Catalog of Visual Double Stars, Publ.~Lick Obs., No.~21
\bibitem[Jerzykiewicz(1999)]{jerz99} Jerzykiewicz, M. 1999, \nar, 43, 455
\bibitem[Jerzykiewicz(2007)]{jerz07} Jerzykiewicz, M. 2007, \actaa, 57, 33
\bibitem[Jerzykiewicz \& Pigulski(1996)]{jepi96} Jerzykiewicz, M., \& Pigulski, A. 1996, \mnras, 282, 853
\bibitem[Jerzykiewicz \& Pigulski(1998)]{jepi98} Jerzykiewicz, M., \& Pigulski, A. 1998, PASPC, 135, 43
\bibitem[Jerzykiewicz \& Pigulski(1999)]{jepi99} Jerzykiewicz, M., \& Pigulski, A. 1999, \mnras, 310, 804
\bibitem[Jerzykiewicz \& Sterken(1992)]{jest92} Jerzykiewicz, M., \& Sterken, C. 1992, \mnras, 257, 303
\bibitem[Jerzykiewicz et al.(1988)]{jerz88} Jerzykiewicz, M., Sterken, C., \& Kubiak, M. 1988, \aaps, 72, 449
\bibitem[Kazarovets \& Samus(1997)]{kasa97} Kazarovets, E.V., \& Samus, N.N. 1997, IBVS, 4471
\bibitem[Keenan(1986)]{keen86} Keenan, F.P. 1986, Irish A.\,J., 17, 483
\bibitem[Keenan(1992)]{keen92} Keenan, F.P. 1992, \qjras, 33, 325
\bibitem[Keenan et al.(1986)]{klbd86} Keenan, F.P., Lennon, D.J., Brown, P.F.J., \& Dufton, P.L. 1986, \apj, 307, 694
\bibitem[Kilkenny \& Hill(1975)]{kihi75} Kilkenny, D., \& Hill, P.W. 1975, \mnras, 172, 649
\bibitem[Kilkenny \& van Wyk(1990)]{kivw90} Kilkenny, D., \& van Wyk, F. 1990, \mnras, 244, 727
\bibitem[Kilkenny et al.(1975)Kilkenny, Hill \& Schmidt-Kaler]{kilk75} Kilkenny, D., Hill, P.W., \& Schmidt-Kaler, T. 1975, \mnras, 171, 353
\bibitem[Kilkenny et al.(1977)Kilkenny, Hill \& Brown]{kilk77} Kilkenny, D., Hill, P.W., \& Brown, A. 1977, \mnras, 178, 123
\bibitem[Koen(1993)]{koen93} Koen, C. 1993, \mnras, 264, 165
\bibitem[Lampens(1988)]{lamp88} Lampens, P. 1988, Ph.D.~thesis, Katholieke Univ. Leuven
\bibitem[L\'epine et al.(2001)]{lepi01} L\'epine, S., Wallace, D., Shara, M.M., Moffat, A.F.J., \& Niemela, V.S. 2001, \aj, 122, 3407
\bibitem[Lesh \& Aizenman(1974)]{leai74} Lesh, J.R., \& Aizenman M.L. 1974, \aap, 34, 203
\bibitem[Lomb(1978)]{lomb78} Lomb, N.R. 1978, \mnras, 185, 325
\bibitem[Lynn et al.(2002)]{lynn02} Lynn, B.B., Dufton, P.L., Keenan, F.P., et al. 2002, \mnras, 336, 1287
\bibitem[Massey et al.(1981)Massey, Conti \& Niemela]{mass81} Massey, P., Conti, P.S., \& Niemela, V.S. 1981, \apj, 246, 145
\bibitem[Molenda-\.Zakowicz \& Po{\l}ubek(2005)]{mopo05} Molenda-\.Zakowicz, J., \& Po{\l}ubek, G. 2005, \actaa, 55, 375
\bibitem[Monderen et al.(1988)]{mond88} Monderen, P., De Loore, C.W.H., van der Hucht, K.A., \& van Genderen, A.M. 1988, \aap, 195, 179
\bibitem[Morgan et al.(1953)Morgan, Whitford \& Code]{morg53} Morgan, W.W., Whitford, A.E., \& Code, A.D. 1953, \apj, 118, 318
\bibitem[Moskalik \& Dziembowski(1992)]{modz92} Moskalik, P., \& Dziembowski, W.A. 1992, \aap, 256, L5
\bibitem[Nesterov et al.(1995)]{nest95} Nesterov, V.V., Kuzmin, A.V., Ashimbaeva, N.T., et al. 1995, \aaps, 110, 367
\bibitem[Niemela et al.(1998)]{niem98} Niemela, V.S., Shara, M.M., Wallace, D.J., Zurek, D.R., \& Moffat, A.F.J. 1998, \aj, 115, 2047
\bibitem[Paardekooper et al.(2002)]{paar02} Paardekooper, S.J., Veen, P.M., Van Genderen, A.M., \& van der Hucht, K.A. 2002, \aap, 384, 1012
\bibitem[Pamyatnykh(1999)]{pamy99} Pamyatnykh, A.A. 1999, \actaa, 49, 119
\bibitem[Pigulski(1992)]{pigu92} Pigulski, A. 1992, \aap, 261, 203
\bibitem[Pigulski(1993)]{pigu93} Pigulski, A. 1993, \aap, 274, 269
\bibitem[Pigulski(1994)]{pigu94} Pigulski, A. 1994, \aap, 292, 183
\bibitem[Pigulski(2005)]{pigu05} Pigulski, A. 2005, \actaa, 55, 219 (P05)
\bibitem[Pigulski \& Boratyn(1992)]{pibo92} Pigulski, A., \& Boratyn, D.A., 1992, \aap, 253, 178
\bibitem[Pigulski \& Pojma\'nski(2007)]{pipo07} Pigulski, A., \& Pojma\'nski, G., 2007, \aap, submitted (Paper II)
\bibitem[Pojma\'nski(1997)]{pojm97} Pojma\'nski, G. 1997, \actaa, 47, 467
\bibitem[Pojma\'nski(1998)]{pojm98} Pojma\'nski, G. 1998, \actaa, 48, 35
\bibitem[Pojma\'nski(2000)]{pojm00} Pojma\'nski, G. 2000, \actaa, 50, 177
\bibitem[Poj\-ma\'n\-ski(2001)]{pojm01} Pojma\'nski, G. 2001, PASPC, 246, 53
\bibitem[Pojma\'nski(2002)]{asas1} Pojma\'nski, G. 2002, \actaa, 52, 397
\bibitem[Pojma\'nski(2003)]{asas2} Pojma\'nski, G. 2003, \actaa, 53, 341
\bibitem[Pojma\'nski \& Maciejewski(2004)]{asas3} Pojma\'nski, G., \& Maciejewski, G. 2004, \actaa, 54, 153
\bibitem[Pojma\'nski \& Maciejewski(2005)]{asas4} Pojma\'nski, G., \& Maciejewski, G. 2005, \actaa, 55, 97
\bibitem[Pojma\'nski et al.(2005)Pojma\'nski, Pilecki \& Szczygie{\l}]{asas5} Pojma\'nski, G., Pilecki, B., \& Szczygie{\l}, D. 2005, \actaa, 55, 275
\bibitem[Ramspeck et al.(2001)Ramspeck, Heber \& Moehler]{rams01} Ramspeck, M., Heber, U., \& Moehler, S. 2001, \aap, 378, 907
\bibitem[Roberts(1962)]{robe62} Roberts, M.S. 1962, \aj, 67, 79
\bibitem[Rufener(1981)]{rufe81} Rufener, F. 1981, \aaps, 45, 207
\bibitem[Rufener \& Bartholdi(1982)]{ruba82} Rufener, F., \& Bartholdi, P. 1982, \aaps, 48, 503
\bibitem[Ryans et al.(1996)]{ryan96} Ryans, R.S.I., Hambly, R.C., Dufton, P.L., \& Keenan, F.P. 1996, \mnras, 278, 132
\bibitem[Schild(1970)]{schi70} Schild, R.E. 1970, \apj, 161, 855
\bibitem[Schrijvers \& Telting(2002)]{scte02} Schrijvers, C., \& Telting, J.H. 2002, \aap, 394, 603
\bibitem[Shobbrook(1979)]{shob79} Shobbrook, R.R. 1979, \mnras, 189, 571
\bibitem[Shobbrook(1984)]{shob84} Shobbrook, R.R. 1984, \mnras, 206, 273
\bibitem[Smith(1968)]{smit68} Smith, L.F. 1968, \mnras, 141, 317
\bibitem[Stankov \& Handler(2005)]{stha05} Stankov, A., \& Handler, G. 2005, \aaps, 158, 193
\bibitem[Stankov et al.(2002)]{stan02} Stankov, A., Handler, G., Hempel, M., \& Mittermayer, P. 2002, \mnras, 336, 189
\bibitem[Sterken \& Jerzykiewicz(1990)]{stje90} Sterken, C., \& Jerzykiewicz, M. 1990, \actaa, 40, 123
\bibitem[Sterken et al.(1986)Sterken, Jerzykiewicz \& Manfroid]{ster86} Sterken, C., Jerzykiewicz, M., \& Manfroid, J. 1986, \aap, 169, 166
\bibitem[Tobin et al.(1994)Tobin, Viton \& Sivan]{tobi94} Tobin, W., Viton, M., \& Sivan, J.-P. 1994, \aaps, 107, 385
\bibitem[van Genderen et al.(1990)van Genderen, van der Hucht \& Larsen]{vgen90} van Genderen, A.M., van der Hucht, K.A., \& Larsen, I. 1990, \aap, 229, 123
\bibitem[van Genderen et al.(1991)]{vgen91} van Genderen, A.M., Verheijen, M.A.W., van der Hucht, K.A., et al. 1991, IAU Symp.~143, 129
\bibitem[Wa\-el\-kens \& Cuypers(1985)]{wacu85} Waelkens, C., \& Cuypers, J. 1985, \aap, 152, 15
\bibitem[Wa\-el\-kens \& Heynderickx(1989)]{wahe89} Waelkens, C., \& Heynderickx, D. 1989, \aap, 208, 129
\bibitem[Wa\-el\-kens \& Rufener(1983)]{waru83} Waelkens, C., \& Rufener, F. 1983, \aap, 119, 279
\bibitem[Wa\-el\-kens \& Rufener(1988)]{waru88} Waelkens, C., \& Rufener, F. 1988, \aap, 201, L5
\bibitem[Wa\-el\-kens et al.(1991)Waelkens, Van den Abeele \& Van Winckel]{wael91} Waelkens, C., Van den Abeele, K., \& Van Winckel, H. 1991, \aap, 251, 69
\bibitem[Wa\-el\-kens et al.(1998)]{wael98} Waelkens, C., Aerts, C., Kestens, E., Grenon, M., \& Eyer, L. 1998, \aap, 330, 215
\bibitem[Walborn(1971)]{walb71} Walborn, N.R. 1971, \apjs, 23, 257
\bibitem[Whiteoak(1963)]{whit63} Whiteoak, J.B. 1963, \mnras, 125, 105
\end{thebibliography}
\end{document}